\documentclass[preprint,12pt, authoryear]{elsarticle}

\usepackage{amssymb}
\usepackage{amsmath}
\usepackage{geometry}                   		% Allows definition of text/page measurements
\usepackage{multirow}                   		% Multirow and multicolumn spanning with latex tables
\usepackage{enumerate}     
\usepackage{wrapfig}
\usepackage{lscape}
\usepackage{epstopdf}
\usepackage{rotating}                   		% Allows Landscape Orientation in Tables & Figures \begin{sidewasytable}
\usepackage{supertabular}               		% Multiple page tables
\usepackage{lscape}
\usepackage[obeyspaces]{url}            		% line-breaks for long URLs, DOIs, etc                     		% Font
\usepackage{float}
\usepackage{threeparttable} 
\usepackage{graphicx}                   		% Display graphics 
\usepackage{amsfonts}      
\usepackage{hhline}                     			% Works with array package to form single and double lines
\usepackage{longtable}                  		% Produces tables across multiple pages
\usepackage{setspace}       
\usepackage{rotating}                   		% Allows Landscape Orientation in Tables & Figures
\usepackage{multirow}                   		% Multi-row and multi-column spanning latex tables
\usepackage{threeparttable}             		% Three part tables, caption, table, notes
\usepackage{longtable}                  		% Produces tables across multiple pages
\usepackage{tabularx}                           % Tables; column width adjusted to fit page
\usepackage{array}                              % Flexible column formatting; must-have package
\usepackage{booktabs}  
\usepackage{graphicx}
\usepackage[small,bf]{caption}  	
\usepackage{subfig}
\usepackage{gensymb}
\usepackage{ amssymb }
\usepackage{relsize}
\usepackage{adjustbox}
\usepackage{pdflscape}
\journal{Physics of the Earth and Planetary Interiors}

\begin{document}

\begin{frontmatter}



\title{Local Flow Estimation at the top of the Earth's Core using Physics Informed Neural Networks}

\author[inst1]{Naomi Shakespeare-Rees}% \corref{cor1}}

%\cortext[cor1]{Contact Email Address: eenmsr@leeds.ac.uk}
\author[inst1]{Philip W. Livermore}
\author[inst1]{Christopher J. Davies}
\author[inst1]{Hannah F. Rogers}
\author[inst2]{William J. Brown}
\author[inst2]{Ciar\'an D. Beggan}
\author[inst3]{Christopher. C. Finlay}

\affiliation[inst1]{organization={School of Earth and Environment},%Department and Organization
            addressline={University of Leeds, Woodhouse Lane}, 
            city={Leeds},
            postcode={LS2 9JT}, 
            country={UK}}
                       
\affiliation[inst2]{organization={British Geological Survey},%Department and Organization
            addressline={Edinburgh}, 
            country={UK}}

\affiliation[inst3]{organization={DTU Space},%Department and Organization
            addressline={Technical University of Denmark}, 
            city={Kongens Lyngby},
            country={Denmark}}

\begin{abstract}
The Earth’s main geomagnetic field arises from the constant motion of the fluid outer core. By assuming that the field changes are advection-dominated, and that diffusion only plays a minor role, the fluid motion at the core surface can be related to the secular variation of the geomagnetic field, providing an observational approach to understanding the motions in the deep Earth. The majority of existing core flow models are global, showing features such as an eccentric planetary gyre, with some evidence of rapid regional changes. By construction, the flow defined at any location by such a model depends on all magnetic field variations across the entire core-mantle boundary: because of this nonlocal dependence of the flow on the magnetic field, it is very challenging to interpret local structures in the flow as due to specific local changes in magnetic field. Here we present an alternative strategy in which we construct regional flow models that rely only on local secular changes. We use a novel technique based on machine learning termed Physics-Informed Neural Networks (PINNs), in which we seek a regional flow model that simultaneously fits both the local magnetic field variation and dynamical conditions assumed satisfied by the flow. Although we present results using the Tangentially Geostrophic flow constraint, we set out a modelling framework for which the physics constraint can be easily changed by altering a single line of code. After validating the PINN-based method on synthetic flows, we apply our method to the CHAOS-8.1 geomagnetic field model, itself based on data from Swarm. Constructing a global mosaic of regional flows, we reproduce the planetary gyre, providing independent evidence that the strong secular changes at high latitude and in equatorial regions are part of the same global feature. Our models also corroborate regional changes in core flows over the last decade. In our models, we find that the azimuthal flow under South America has changed sign quasi-periodically, with a recent sign change in 2022. Furthermore, our models endorse the existence of a dynamic high latitude jet, which began accelerating around 2005 but has been weakening since 2017. 
\end{abstract}


\end{frontmatter}

\section{Introduction}
\label{sec:intro}
\par
The Earth's main magnetic field, which is generated by a self-sustaining geodynamo arising from fluid motions in the Earth's core \citep{Bullard_1950}, exhibits fluctuations on timescales of seconds to millennia and longer \citep{CONSTABLE2023107090}. Changes in the geomagnetic field on timescales of years to millennia are termed Secular Variation (SV) \citep{JACKSON2015137}. The geomagnetic field has been measured by networks of ground-based observatories since 1837 \citep{Macmillan2007}, and supplemented by continuous satellite measurements since 1999 (\citet{JACKSON2015137}, \citet{Friis-Christensen_Lühr_Hulot_2006}, \citet{201861}, \citet{14-zhangkeke}, \citet{https://doi.org/10.1029/2024GL112305}). Changes in the geomagnetic field can have critical impacts on both industry and scientific exploration in a diverse range of disciplines, such as navigation, satellite operations and the protection of the Earth from space weather. By mapping the outer core flow that generates the SV, the dynamics and properties of the core can be explored, and SV forecasts such as the candidate models of IGRF-13 \citep{IGRF13gen} constructed.

\par
The typical approach to map core flows is to use observations from satellites, observatories and other surveys to construct global geomagnetic field models using spherical harmonics, which can be downward-continued to the Core-Mantle Boundary (CMB). The inversion of the SV to recover motions on the outer core is under-determined, and so additional assumptions must be included. A widely used assumption is that of \textit{frozen flux} \citep{Roberts_Scott_1965}, where on sufficiently short timescales, magnetic diffusion can be considered negligible.  Under this simplification, the magnetic field lines are `frozen' into packets of moving liquid at the surface of the outer core, so that the field becomes a tracer for the flow \citep{Bloxham_Jackson_1991}. However, even with this assumption, the problem of non-uniqueness remains \citep{Backus_1968}. There are still more unknown components of the flow than the number of equations describing them. This means additional constraints on the flow are required in order to reduce the fundamental non-uniqueness. Various flow assumptions have been used in previous studies, including steady flows \citep{doi:10.1098/rsta.1982.0084}, toroidal only flow \citep{Whaler_1980}, tangential geostrophy (TG) \citep{LeMouël1984}, \citep{10.1046/j.1365-246x.2000.00097.x}, quasi-geostrophic flow \citep{10.1111/j.1365-246X.2008.03741.x} and helical and columnar flow \citep{Amit_Olson_2004}. Non-uniqueness is further reduced by the large scale flow assumption, in which small scale flow structures are penalised  \citep[e.g.][]{Bloxham_1988}.   
\par
Spherical harmonics are also used to create these global flow models, and underpin most studies to date. Results from previous global outer core flow inversions, as well as numerical simulations, reveal an eccentric, anticyclonic, planetary gyre \citep{10.1111/j.1365-246X.2008.03741.x}. This gyre is a planetary-scale circulation pattern, travelling west under the Atlantic, then north under Asia, and then westwards near the North Pole. Although this is a global pattern, there are several regional features that are of specific interest. Firstly, there is evidence for a localised jet under the Bering strait, which has been strengthening since 2005 and is associated with a strong change in the SV at high northern latitudes \citep{Livermore_Hollerbach_Finlay_2017}. Additionally, multiple studies, such as \citet{Whaler_Hammer_Finlay_Olsen_2022} and \citet{Li_Lin_Zhang_2024}, have observed changes in the azimuthal flow direction in areas in the equatorial region, particularly beneath Indonesia and Central America. These may be a result of hydromagnetic waves producing changes on interannual timescales \citep{Gillet_Gerick_Jault_Schwaiger_Aubert_Istas_2022}, and many of these sign changes may be associated with geomagnetic jerks - a phenomenon in which the SV changes rapidly, sometimes in a spatially localised manner \citep{Brown_Mound_Livermore_2013}. 
%\subsection{Local Flow Inversions}
\par 
In contrast to global core flow inversions, we adopt a local approach in which flows are inferred from a regional realisation of a global geomagnetic field model, presenting an independent method to check global flow analysis. This is important, as a large-scale global inversion will act to interpolate multiple local features into one larger one. For example, a global approximation to westward drift in the Atlantic and flow into the polar regions would be a gyre that connects these regions, and so the use of regional inversions could test how robust these features are. Additionally, in a global flow analysis, any point on the CMB depends non-locally on all the points in the model, but in a local flow inversion the same point would only depend on the other points in the region. As the data coverage of the Earth is spatially uneven, especially at the poles, a regional methodology would probe the reliability of local features in these areas, ensuring they are not an artefact of a global inversion. This would allow for regional features to be studied in more detail, without the uncertainty of coupling to neighbouring regions. 
\par
Local core flow inversions have been attempted in two previous studies with mixed results. \citet{Rogers_2022} studied regional variations using spherical Slepian functions. This technique produced better separation of SV at the Earth's surface compared to spherical harmonics but reliable local flow separations and SV separations at the CMB were not achieved. \citet{10.1093/gji/ggad089} presented a local core flow inversion methodology based on pointwise inversion, which was able to reproduce the main features found in global core flow studies, but it was found that additional smoothing was required to prevent unreliable re-construction of small-scale flows. They also found that their results heavily depended on what prior they used to reduce the non-uniqueness.
%\subsection{Machine Learning}
%\label{sec:ml}
\par
This work introduces a novel approach to infer local flows, employing recent advancements in machine learning through the use of Physics-Informed Neural Networks (PINNs). Machine learning refers to a set of statistical techniques to leverage data in order to undertake a task, without being explicitly programmed to do so \citep{Armstrong_Fletcher_2019}. This
allows the user to extract knowledge and draw inferences from data \citep{Jordan_Mitchell_2015}. Neural Networks (NN) are a machine learning technique, consisting of layers of artificial neurons that can process information. PINNs, first proposed in \citet{Raissi_Perdikaris_Karniadakis_2019}, are a class of neural network that use mathematical descriptions of physical laws as constraints in order to solve forward and inverse problems. PINNs have been used in a diverse range of fields such as fluid mechanics \citep{Raissi_Yazdani_Karniadakis_2020}, medicine \citep{Arzani_Wang_D’Souza_2021}, nuclear physics \citep{Schiassi_DeFlorio_Ganapol_Picca_Furfaro_2022} and seismology \citep{Chen_deRidder_Rost_Guo_Wu_Chen_2022}. PINNs have been recently used in core flow and length of day (LOD) analysis by \citet{Li_Lin_Zhang_2024}, albeit in a global flow inversion framework, rather than the local approach that we adopt in this study.
\par
The methodology presented in this work takes secular variation from regional latitude-longitude boxes from the global geomagnetic field model CHAOS-8.1 as input, and then outputs a flow that both re-produces the input secular variation and satisfies an additional flow constraint. We aim to establish a framework that can be used for any flow assumption, which we illustrate here using the Tangentially Geostrophic (TG) flow constraint. We do not rely on any prior information other than the physics constraint. The methodology and validation is described in section \ref{sec:method}. Results, including local flow analysis as well as a global flow model constructed from a mosaic of regional models, are presented in section \ref{sec:res}. All of these results are discussed in section \ref{sec:discussion}, with conclusion in section \ref{sec:conc}.

\newpage
\section{Method}
\label{sec:method}
\subsection{Mathematical Framework}
\par
We model the Earth's core as a sphere, described using spherical coordinates ($r, \theta, \phi$), and assume that the flow within it obeys a non-penetration boundary condition at the edge of the core such that $u_r = 0$ at the CMB. 
Assuming frozen flux, the radial magnetic field, $B_r$, can be related to the outer core surface horizontal fluid movement via the radial component of the induction equation, at the core surface, just below the CMB. This is written as
\begin{equation}
\label{eqn:induct}
\frac{\partial B_r}{\partial t} = -\nabla_H \cdot (\pmb{u}B_r),
\end{equation}
in which $ \pmb{u} = [0, u_\theta, u_\phi]$ is the flow, what is what is sought in the inversion methodology, and $\nabla_H$ is the horizontal gradient operator. 

%\subsection{Tangential Geostrophy Constraint}
\par
The tangential geostrophic flow assumption arises from considering the force balances at the top of the Earth's core \citep{LeMouël1984}. If the force balance is dominated by the Coriolis (rotational) and pressure forces, then the Navier-Stokes equation for the motion of the fluid reduces to:
\begin{equation}
\label{eqn:forcebal}
2 \rho (\pmb{\Omega} \times \pmb{u})_H + \nabla_H p = 0,
\end{equation}
where $\rho$ is the density of the liquid outer core, $\pmb{\Omega}$ is the rotation vector of the Earth, and $p$ is the pressure of the fluid. 
\par
Taking the horizontal divergence of the cross product of equation \eqref{eqn:forcebal} and the radial unit vector $\hat{\pmb{r}}$ gives
\begin{equation}
\label{eqn:tg_long}
\nabla_H \cdot (\pmb{u} \cos{\theta}) =0,
\end{equation}
which can be written as
\begin{equation}
\label{eqn:tg}
\frac{\tan{\theta}}{r}u_{\theta} - \nabla_H \cdot \pmb{u}  =0,
\end{equation}
that defines the TG constraint (\citet{LeMouël1984}, \citet{Amit_Pais_2013}). This additional constraint modifies equation \eqref{eqn:induct} to
\begin{equation}
\label{eqn:tginduct}
\frac{\partial B_r}{\partial t} = -\pmb{u}\cos{\theta} \cdot \nabla_H \left(\frac{B_r}{\cos{\theta}}\right).
\end{equation}
It was noted by \citet{Backus_LeMouël_1986} that this flow is non-unique on contours of $B_r/\cos{\theta}$, as well as in regions bounded by contours of $B_r/\cos{\theta}$ that do not cross the equator, meaning care must be taken when interpreting flows in these areas. 
\par
In spherical coordinates, the total flow $\bf{u}$ can be decomposed into  the toroidal and poloidal flows
\begin{equation} {\pmb{u}} = \nabla \times T(\theta,\phi) {\bf r}  + \nabla_H (r P(\theta,\phi)),
\end{equation}
whose horizontal components can be written \citep{HOLME201591}, 
\begin{equation}
\label{eqn:u_t}
\pmb{u}_T = \left(\frac{1}{\sin{\theta}}\frac{\partial \textit{T}}{\partial \phi}, -\frac{\partial\textit{T}}{\partial\theta}\right),
\end{equation}
\begin{equation}
\label{eqn:u_p}
\pmb{u}_P = \left(\frac{\partial\textit{P}}{\partial \theta}, \frac{1}{\sin{\theta}}\frac{\partial\textit{P}}{\partial\phi}\right).
\end{equation}
Therefore, in our flow inversion we seek toroidal ($T$) and poloidal ($P$) scalar functions which define the flow. 
\subsection{Neural Networks}
\begin{figure}[H]
    \centering
    \captionsetup{justification=centering,margin=0.25cm}
    \includegraphics[scale = 0.09]{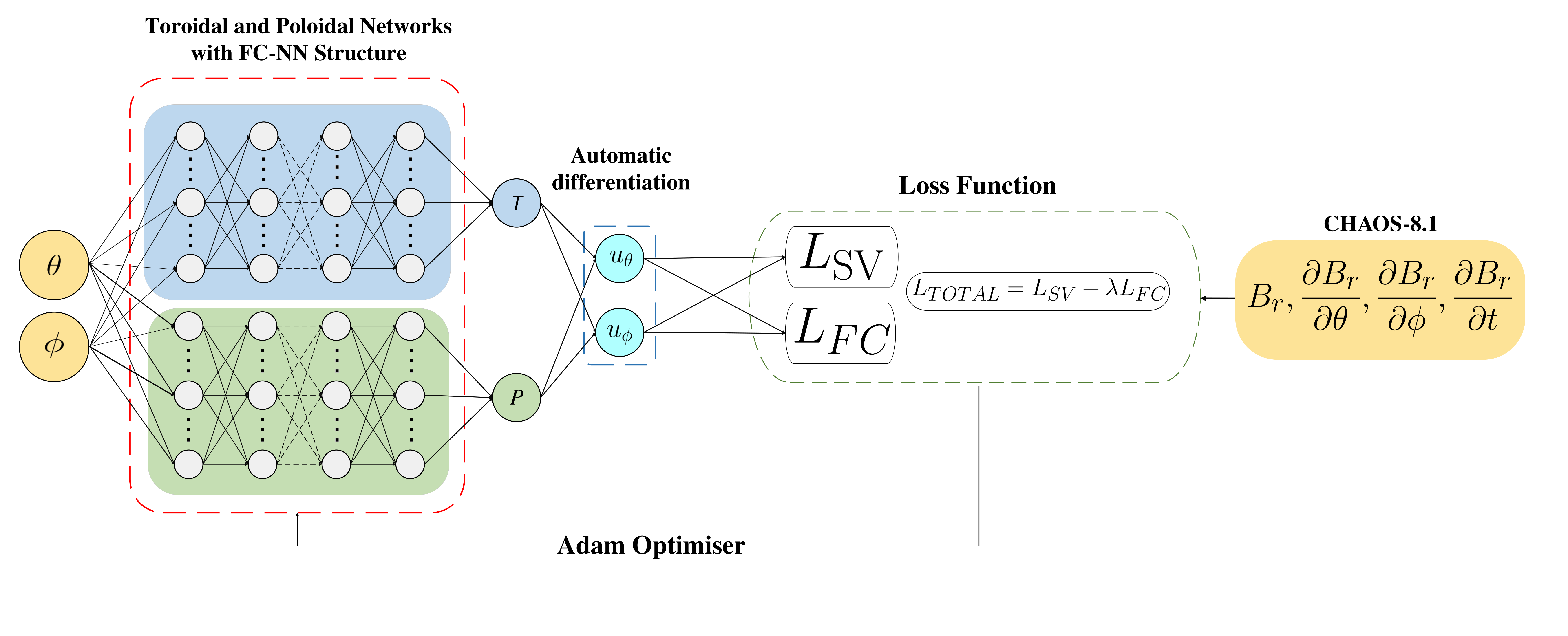}
    \caption{Schematic of the PINN used in this study, where $\theta, \phi$ are co-latitude and longitude, $T$ and $P$ are the toroidal and poloidal scalars, and  $u_\theta$ and $u_\phi$  are the components of the horizontal flow. Orange indicates inputs to the network, the two parallel FC-NNs are blue and green, loss functions in white, and the outputs are cyan.} 
    \label{fig:schematic}
\end{figure}

Our inversion methodology consists of two Fully Connected Neural Networks (FC-NNs) working in parallel: one to describe the toroidal scalar \textit{T} and the other to describe the poloidal scalar \textit{P}. A schematic for this is shown in figure \ref{fig:schematic}. FC-NNs consist of layers of nodes, organised into an input layer, multiple hidden layers, and an output layer \citep{Jordan_Mitchell_2015}.
Each of these nodes are connected to each other and have an associated weight and bias, and the output of each node is computed by some non-linear function of the input and the weight. This non-linear function is called the activation function, and the weights and biases are adjusted during a process termed training \citep{Goodfellow-et-al-2016}. Once the network weights are determined, the flows are then described using equations \eqref{eqn:u_t} and \eqref{eqn:u_p}. 
\par
While we could train a network without any flow assumptions, here we use an additional physics constraint to enforce TG. This takes place solely in the training stage, which is done by minimising loss term consisting of a data loss, which aims to implement equation \eqref{eqn:induct}, and a physics-based loss, which aims to implement equation \eqref{eqn:tg}. In this way, we fit both to the data and to the underlying physics. This is so we can establish a framework wherein different flow assumptions, such as Toroidal or Helical flow, can be swapped in and out using a single line of code. All the loss terms are implemented by defining a derived quantity from the network and a corresponding target, with the loss value determined by the Mean Square Error (MSE) between the quantity and the target. The total loss, $L_{TOTAL}$ is defined by
\begin{equation}
\label{eqn:loss}
    \begin{aligned}
    L_\textrm{TOTAL}=L_\textrm{SV}+\lambda L_\textrm{FC},
    \end{aligned}
\end{equation}
which consists of two loss terms:
\begin{itemize}
    \item The Data Loss ($L_\textrm{SV})$:
    \begin{equation}
    \label{eqn:svloss} 
        \begin{aligned}
         L_\textrm{SV} = \frac{1}{N} \mathlarger{\mathlarger{\sum}}_{i}\left[ \left(-\frac{1}{r} \left(\frac{u_\phi}{\sin\theta} \frac{\partial B_r}{\partial \phi} + u_\theta \frac{\partial B_r}{\partial \theta}\right) + B_{r}(\nabla_{H} \cdot \pmb{u}) \right) - \frac{\partial B_r}{\partial t}_{CHAOS}\right]^{2} \bigg\rvert_{\theta_i, \phi_i}
        \end{aligned}
    \end{equation}
    where $N$ is the number of points, and
    \item the TG Flow Loss ($L_\textrm{FC}$): 
    \begin{equation}
        \label{eqn:fcloss} 
            \begin{aligned}
             L_\textrm{FC} =  \frac{1}{N}\mathlarger{\mathlarger{\sum}}_{i} \left[\nabla_{H} \cdot \pmb{u} -\left(\frac{\tan\theta}{r}\right) u_{\theta}\right]^2 \bigg\rvert_{\theta_i, \phi_i}
            \end{aligned}
    \end{equation}
\end{itemize}

where $\theta_i, \phi_i$ are the pre-determined set of latitude-longitude grid points on which we impose the constraints. We use the same grid to constrain the data and the physics. For the data constraint, the target is the SV from CHAOS-8.1, whereas in the flow constraint the target is zero. A scaling analysis of typical magnitudes of the two terms suggests a value of $\lambda = 1000$ for the weighting factor, which is applied to $L_{FC}$ in equation \eqref{eqn:loss} so that both terms are $\mathcal{O}(1)$. This is desirable so that one loss term is not more important than the other.
\par
The networks are trained using the optimisation method Adam \citep{Kingma_Ba_2017}, which updates both networks simultaneously. Training the network occurs in two steps: forward-propagation and back-propagation. In forward-propagation, the PINN takes in a ($\theta, \phi$) grid of points, at a radius of 3485 km, and then analytically computes the toroidal and poloidal scalar value at each of those points, as well as the derivatives. These derivatives are then used in equations \ref{eqn:u_t} and \ref{eqn:u_p} to produce $u_\theta, u_\phi$. The total error is calculated by summing $L_{SV}$, the error between the network derived SV and the CHAOS-8.1 SV, and $L_{FC}$, the departure of the flow from the tangentially geostrophic flow assumption, together with a weighting factor $\lambda$. This error is then back-propagated through the network by the Adam optimiser, adjusting the weights and biases of the network to attempt to minimise the MSE error. This process repeats for 100,000 iterations until the loss is minimised such that the loss does not decrease further. The PINN methodology is written with PyTorch Version 2.5 \citep{paszke2019pytorchimperativestylehighperformance}. Our dataset is of order 1000-10000 points, which is modest by machine learning standards, and so we are able to use all the data in each iteration of training. When training the network, we initialise the weights from a psuedo-random distribution using a technique called Xavier initialisation \citep{pmlr-v9-glorot10a}. The optimiser then implements a descent method to seek a global minimum. However, we found empirically that $L_{TOTAL}$ as a function of the model parameters is very complex, and so the optimiser often only finds local minima. We therefore choose multiple different seeds -- and so initial weights -- for independent training, from which we select the model that achieves the lowest loss. 

%\newpage
\subsection{Training Dataset: CHAOS-8.1}
\par
Our training procedure requires values for the SV to train the PINN, which enters through $L_{SV}$. We use the model CHAOS-8.1 \citep{Kloss_Finlay_Olsen_Tøffner-Clausen_Gillet_Grayver_2024} projected onto a grid in spherical coordinates. CHAOS-8.1 spans from 1999 to 2025, and is built from satellite and ground based observatory data. It is the latest update of the CHAOS family of models, which have been used for multiple core flow inversion studies (eg. \citet{Gillet_Gerick_Jault_Schwaiger_Aubert_Istas_2022}) making it a suitable choice of model for this study. The time-dependent internal field model is defined up to spherical harmonic degree 20, but only the coefficients for the main field and SV up to degree 13 are used here, as at higher degrees the crustal field dominates the magnetic field signal \citep{https://doi.org/10.1029/GL009i004p00250}. CHAOS-8.1 is also temporally regularised in order to reduce non-uniqueness, by taking temporal covariances from geodynamo simulations. This is a departure from the regularisation method of CHAOS-7 \citep{Finlay_Kloss_Olsen_Hammer_Tøffner-Clausen_Grayver_Kuvshinov_2020}, which penalised the second time derivative at the endpoints, and the third time derivative throughout. This difference in regularisation method affects the small scale features of the SV and the intensity of the acceleration, which may have an effect on the recovered flows \citep{Kloss_Finlay_Olsen_Tøffner-Clausen_Gillet_Grayver_2024}.
%Figure \ref{fig:chaos} shows the SV taken from CHAOS-8.1 in 2024, truncated at degree 13. 
%\begin{figure}[H]
%    \centering
%    \captionsetup{justification=centering,margin=0.25cm}
%    \includegraphics[scale = 0.35]{sv.png}
%    \caption{SV from CHAOS-8.1, at epoch 2024.0 \citep{Kloss_Finlay_Olsen_Tøffner-Clausen_Gillet_Grayver_2024}, truncated at degree 13, with continents shown for reference.} 
%    \label{fig:chaos}
%\end{figure}
\par
\begin{figure}[H]
    \centering
    \captionsetup{justification=centering,margin=0.25cm}
    \includegraphics[scale = 0.35]{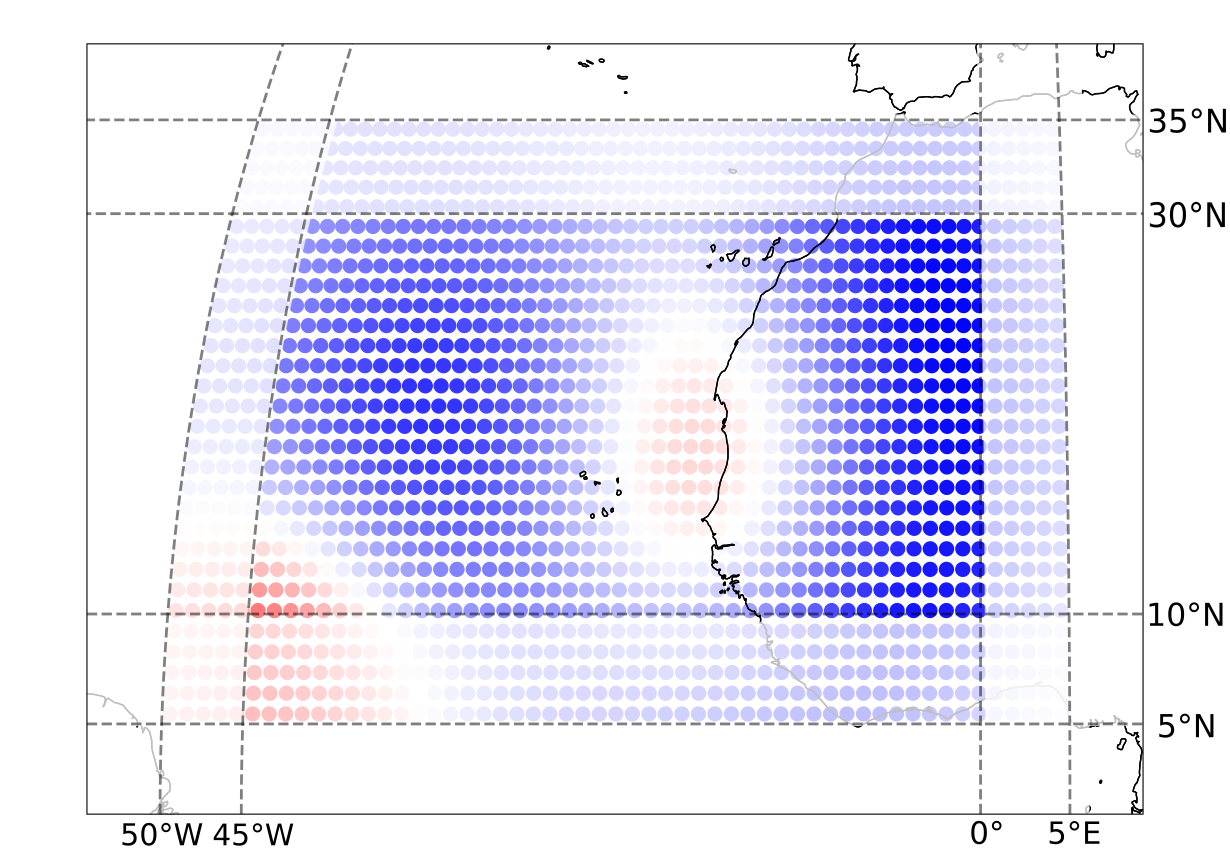}
    \caption{Example of a latitude-longitude box used for inversion. After training, a 5$\degree$ border shown in translucent colouring, is removed. Colours show the radial SV at the CMB from CHAOS-8.1 for reference.} 
    \label{fig:box}
\end{figure}

In order to use CHAOS-8.1 as input to the local flow inversion, a subsection of the SV is selected, either in a latitude-longitude box (as in figure \ref{fig:box}), or in a longitudinal band, each with a grid density of one point per degree. A 5$^\circ$ border is added to the area of interest on all edges, which is removed after training. This is to promote continuity between adjacent regions, because continuity is not explicitly imposed, as well as to avoid edge effects. To maximise the performance of the model, and to minimise training time, the CHAOS-8.1 input and outputs of the network are re-scaled so that they all have a magnitude of about 1 \citep{Sola_Sevilla_1997}.  We do this by measuring $B_r$ in $\mu T$, time in $0.1 years$, length in $km$. Typical values of $B_r$ at the CMB are 500$\mu T$ and so typical values of SV are then 1 $\mu T/0.1 yr$ in these units and flow speeds are typically $1 km/0.1 years$.

\subsection{Parameter Choices}
\label{sec:son}
\par
Having described the algorithm, there remain several parameter choices that define both the model and how it is trained. First, is the learning rate of the model, which governs how rapidly the weights change with each iteration. After testing a variety of learning rates, we found empirically that a standard learning rate of $10^{-3}$ worked well. Second is the choice of activation function. We found that the hyperbolic tangent function worked well, which is widely used due to its zero-centred property that results in faster training \citep{Goodfellow-et-al-2016}. Lastly, we must select a network size - the number of hidden layers and the number of neurons per layer. A small network has a very limited functional representation, and in our case would result in a flow too smooth to fit the SV data. In contrast, a large network can represent spatially complex behaviour, and could result in flows that fit the SV but have spurious small-scale features. We seek to adopt an optimal network size that is large enough to fit the data constraints, but no larger, which effectively penalises small-scale features of the flow. This approach is analogous to the spatial regularisation of global flows, for which their complexity is penalised through an explicit trade-off between data fit and complexity (for example, the strong norm presented in \citet{Bloxham_1988}). To be clear, we do not impose explicit regularisation of our flows, although we could by introducing an additional loss term in equation \eqref{eqn:loss}. 
\par
Our aim is to find a single network size that we can use in each region that we consider. To find the optimum network, we test a variety of network sizes for areas underneath the Atlantic (10$\degree$N to 20$\degree$N latitude, 45$\degree$W to 5$\degree$E longitude) and the South China Sea (10$\degree$N to 20$\degree$N latitude, 100$\degree$E to 130$\degree$E longitude), which represent end member behaviour. Under the Atlantic we expect simple westward flow \citep{Holme_2007}, whereas under the South China Sea we expect complex, diverging flow \citep{Whaler_Hammer_Finlay_Olsen_2022}. For each tested network, we record the SV Root Mean Square Error (RMSE) and the complexity as measured by the average of the squared second spatial derivative of the flow \citep{Bloxham_1988}, evaluated on the SV data grid. We repeat this with 5 different seeds to investigate the spread in results. 
\par
\begin{figure}[H]
    \centering
    \captionsetup{justification=centering,margin=0.25cm}
    \includegraphics[scale = 0.32]{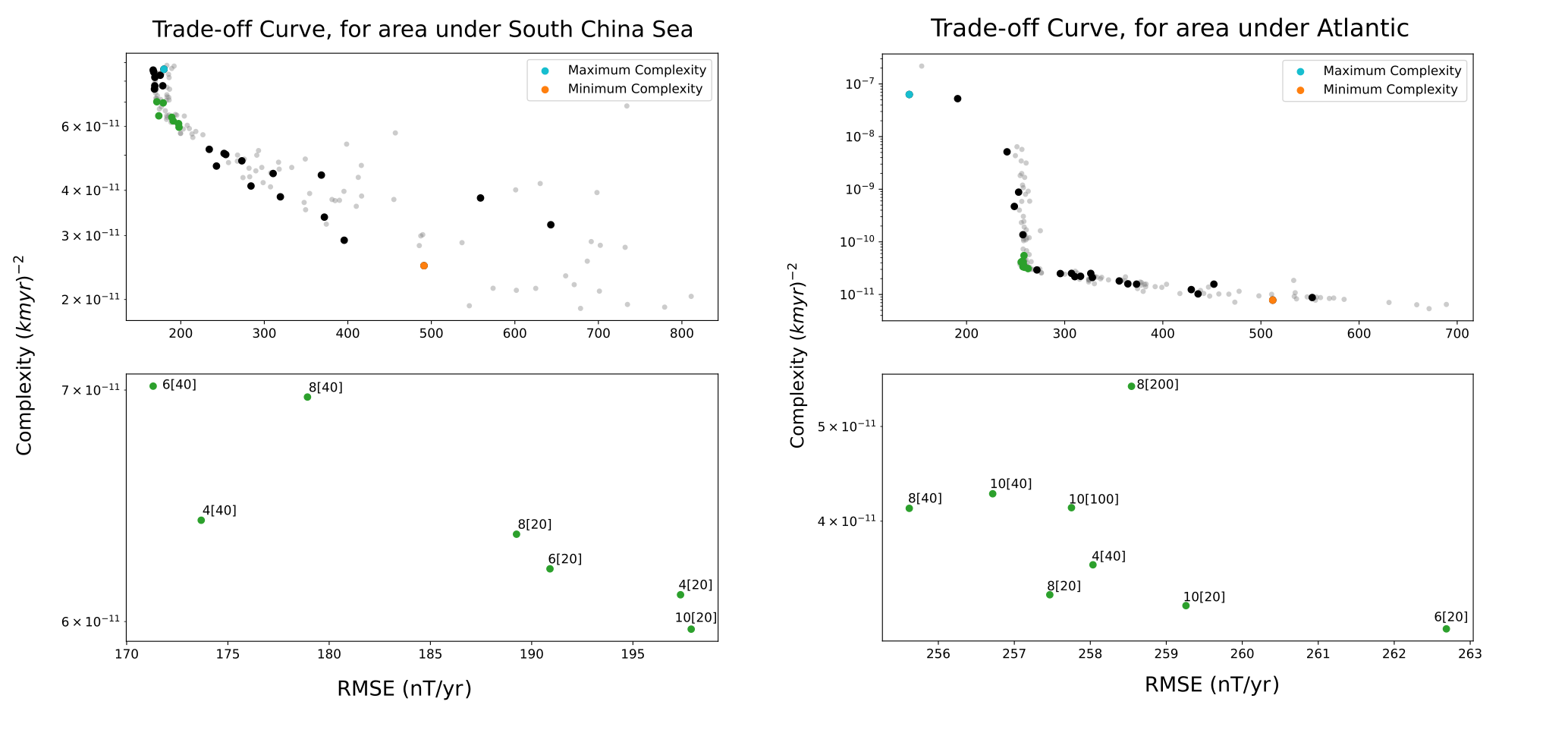}
    \caption{Trade-off curves for an area under the South China Sea (10$\degree$N to 20$\degree$N latitude, 100$\degree$E to 130$\degree$E longitude) (left) and Atlantic (10$\degree$N to 20$\degree$N latitude, 45$\degree$W to 5$\degree$E longitude) (right). Black dots indicate the minimum RMSE for each size of network, grey dots indicate the local minima, and the green dots indicate the points at the knee of the curve, which are shown in more detail on the bottom panel. The labels indicate the size of network, in the format $a[b]$, where $a$ is the number of layers and $b$ is the number of neurons per layer.} 
    \label{fig:to}
\end{figure}
\begin{figure}[H]
    \centering
    \captionsetup{justification=centering,margin=0.25cm}
    \includegraphics[scale = 0.21]{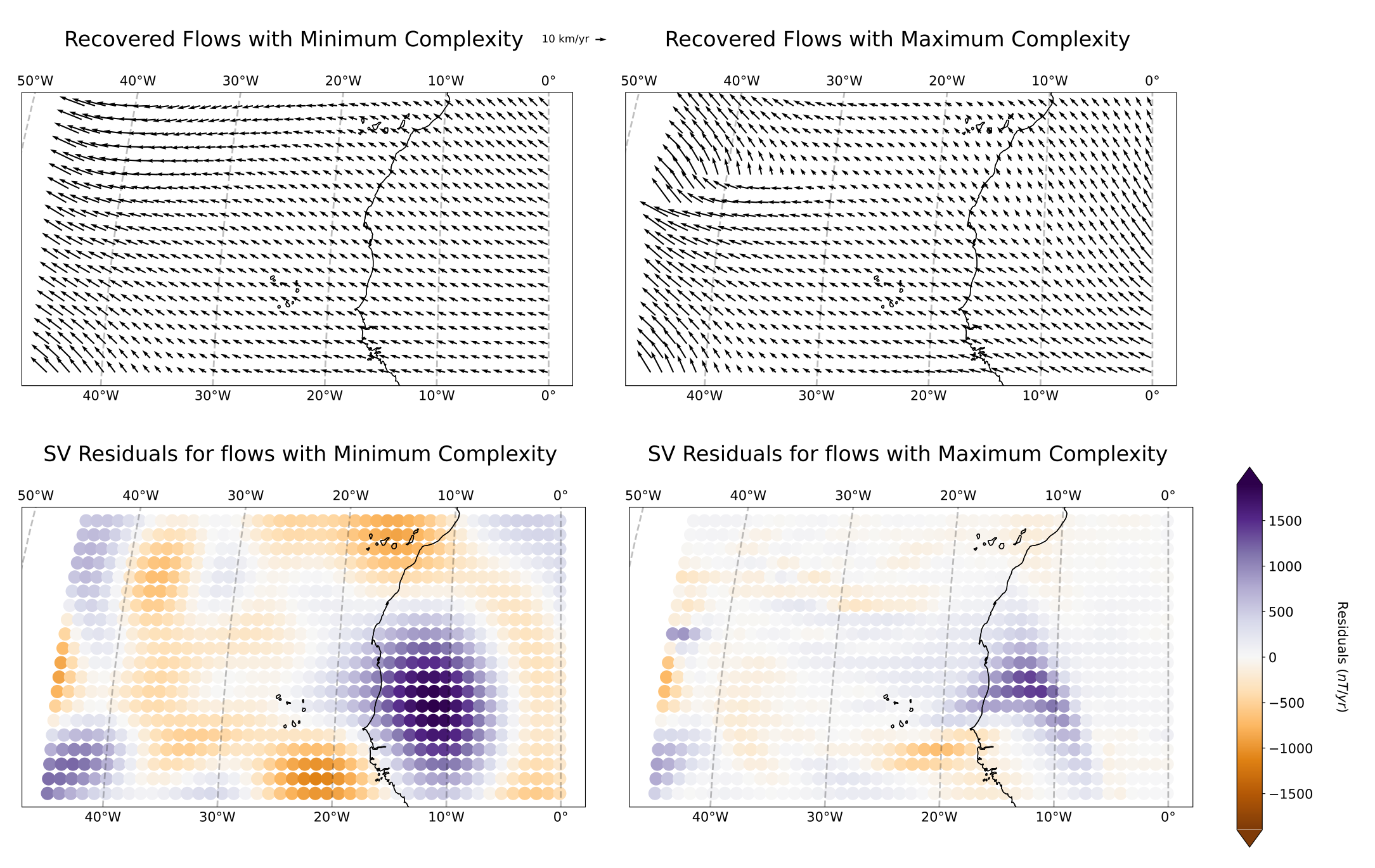}
    \caption{Recovered flows and SV Residuals for the network sizes that produce the least complex (left) and most complex (right) flows, in the area under the Atlantic, at 35$\degree$N to 5$\degree$N latitude.} 
    \label{fig:west_all}
\end{figure}
\begin{figure}[H]
    \centering
    \captionsetup{justification=centering,margin=0.25cm}
    \includegraphics[scale = 0.21]{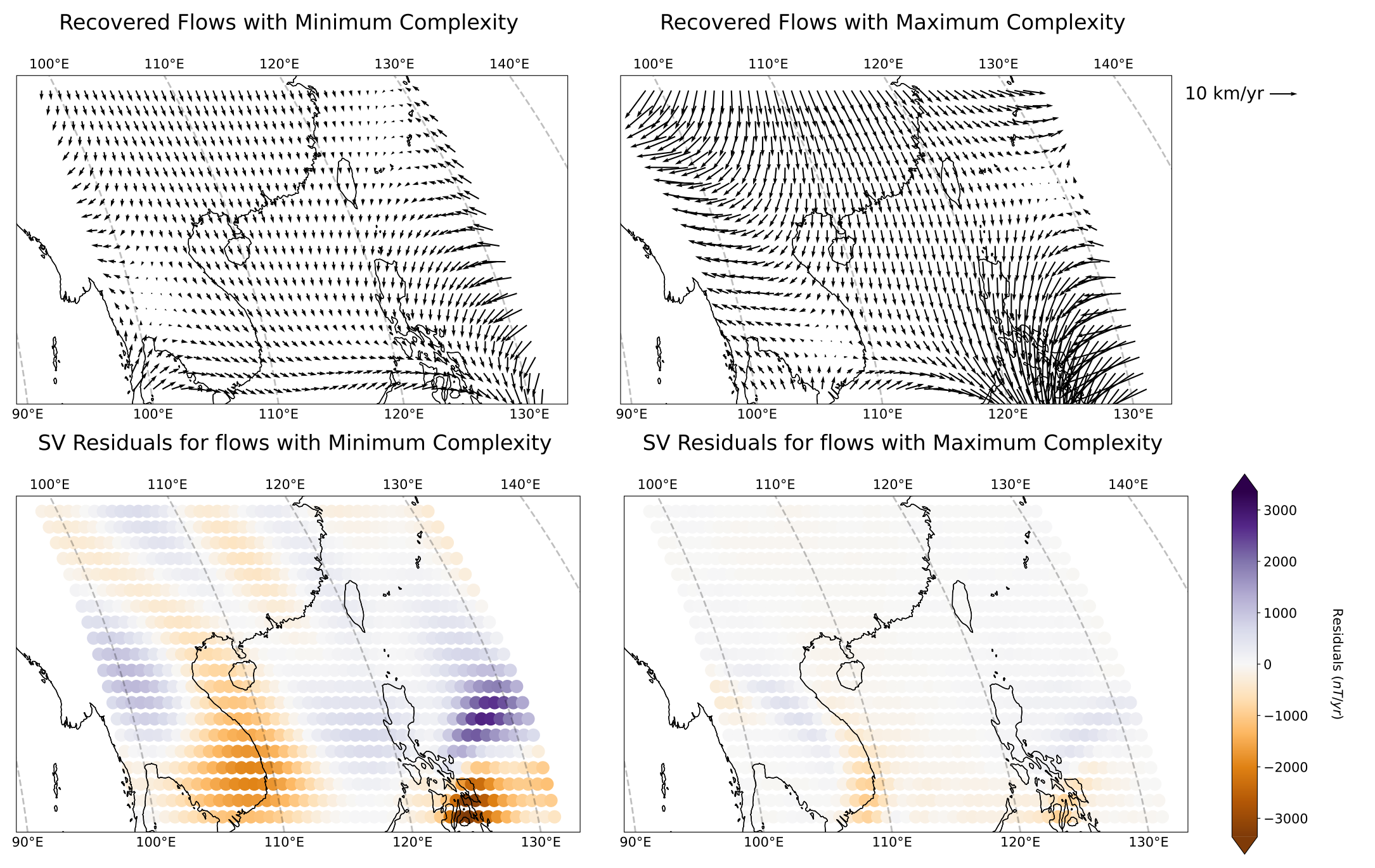}
    \caption{Recovered flows and SV Residuals for the network sizes that produce the least complex (left) and most complex (right) flows, in the area under the South China Sea.} 
    \label{fig:ind_all}
\end{figure}

The results for both regions are plotted in figure \ref{fig:to}, which shows trade-off curves of complexity against RMSE. The grey points show the local minima for all the different seeds, and the black points indicate the seed for each network size which has the minimum RMSE, which we adopt as the closest model we have obtained to the to the global minimum. Simple networks are on the right, as they have a low complexity but a high SV RMSE. Complex networks are on the left, as they have a low SV RMSE but a high complexity. The network size with the highest complexity is shown in blue, whereas the one with the lowest complexity is shown in orange. Examples of the recovered flows and SV residuals at these points are shown in figure \ref{fig:west_all} for the area under the Atlantic, and figure \ref{fig:ind_all} for the area under the South China Sea. For both of these figures, the size of network that recovers the most complex flows fits the data better than the size that recovers the least complex flows, but with the unwanted side effect of adding very rapid, small scale flows. This is due to the non-uniqueness present, both the large and small scale flows fit the SV adequately. An objective way to choose the size of network is to choose the one at the `knee' of this trade-off curve, and the values around this point are shown in green. The bottom panels of figure \ref{fig:to} shows the points at the knee in more detail. The behaviour is not always monotonic, although there is a general trend, perhaps due to the training getting stuck at a local minimum, rather than a global one. We choose 8 layers of 40 neurons, indicated as 8[40] in figure \ref{fig:to}, which was consistently at the knee across regions.

\par
In principle large networks will over-fit the data, which in this case would be evident through high complexities and spurious flows. We note that while the complexity varies across four orders of magnitude in the area under the Atlantic (shown in figure \ref{fig:west_all}), the complexity is never particularly high, with the highest complexity being comparable to those found in \citet{Bloxham_1988}. This is even more apparent under the South China Sea (shown in figure \ref{fig:ind_all}), where the complexity only varies over one order of magnitude. This may be a consequence of using CHAOS-8.1 for the input field and SV, as this model is already smoothed spatially and temporally, and we find that it is apparently difficult to overfit the data in this case. Adding artificial Gaussian noise to the SV data increased the range of complexity magnitudes, and so perhaps smoothing through the choice of the size of networks will be more critical when using `noisier' data sources, such as Geomagnetic Virtual Observatories (GVOs) \citep{Mandea_Olsen_2006}, rather than field models whose complexity was already controlled during their construction.
\subsection{Synthetic Tests}
\label{sec:syn}
\par
In order to test the performance of the chosen networks, a number of synthetic examples were set up with pre-defined flows for the network to discover. On a latitude-longitude grid spanning  -15$^\circ$ to 15$^\circ$ latitude and 0$^\circ$  to 55$^\circ$ longitude with 1-degree spacing in each direction, we define $B_r$ using
\begin{equation}
\label{eqn:18}
B_r = \exp{(-\frac{(\theta - \frac{90\pi}{180})^2 + (\phi - \frac{30\pi}{180})^2}{ 2})},
\end{equation}
which describes a single peak centred at (90$\degree$, 30$\degree$) in radians. For each flow, the instantaneous SV was determined using equation \eqref{eqn:induct}. A range of synthetic flows were investigated, as listed in table \ref{tab:syn}, which span behaviours from simple drift to a more complex vortex. The final synthetic flow tested was a circular Rankine vortex patch first presented in \citet{rankine1872manual}, and most recently used in a geomagnetism context by \citet{AMIT2014110}. The description of this flow does not fit in the table, so it is described here. Rankine flow contains an inner circular area, described by a boundary of angular distance $H$, wherein the fluid motion follows that of solid body rotation with vorticity $\zeta$, and an outer area where the flow does not follow solid body rotation. We only use this inner section for our synthetic flow. The azimuthal toroidal flow, $u_t$, at a given angle from the centre, $h$, is given by
\begin{equation}
\label{eqn:rankine} 
\begin{aligned}
u_t  = -\begin{cases}
    \frac{1}{2} \zeta h, & \text{if $h<H$}\\
    \frac{1}{2} \zeta \frac{H^2}{h^2}, & \text{if $h>H$},
  \end{cases}
\end{aligned}
\end{equation}
where $H = 35^\circ$ is the angular distance of the solid body circular area and $\zeta = 0.1km/0.1year$ is a constant vorticity in the region where $h<H$. The maximum flow speeds are 2.6$km/0.1year$.
\par
\begin{figure}[H]
    \centering
    \captionsetup{justification=centering,margin=0.25cm}
    \includegraphics[scale = 0.3]{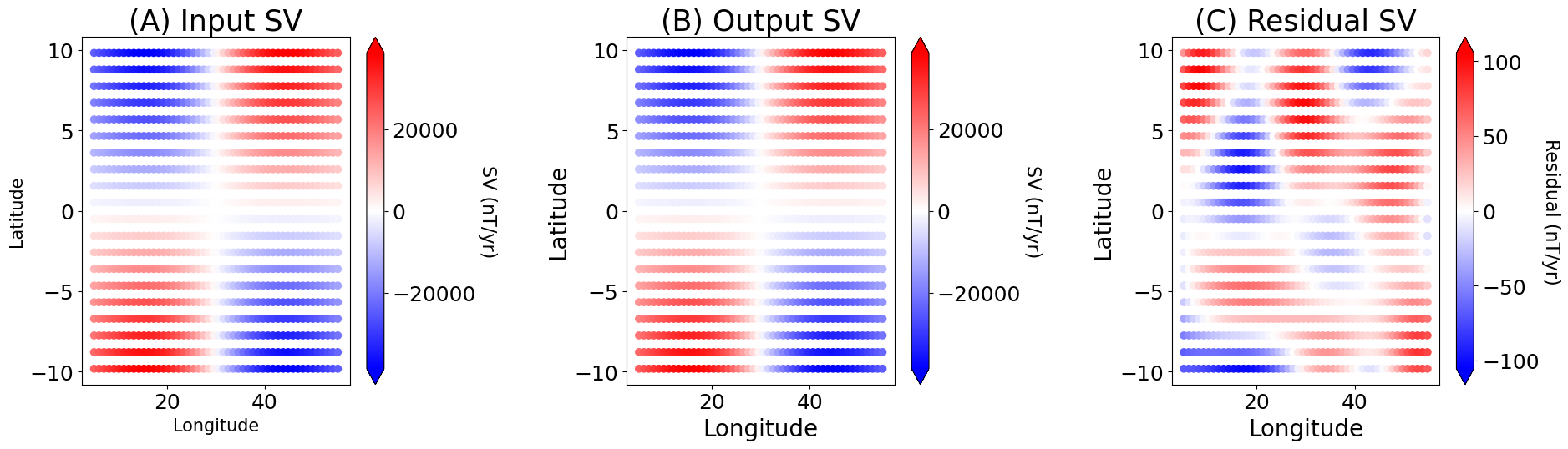}
    \captionsetup{justification=centering,margin=0.25cm}
    \includegraphics[scale = 0.2]{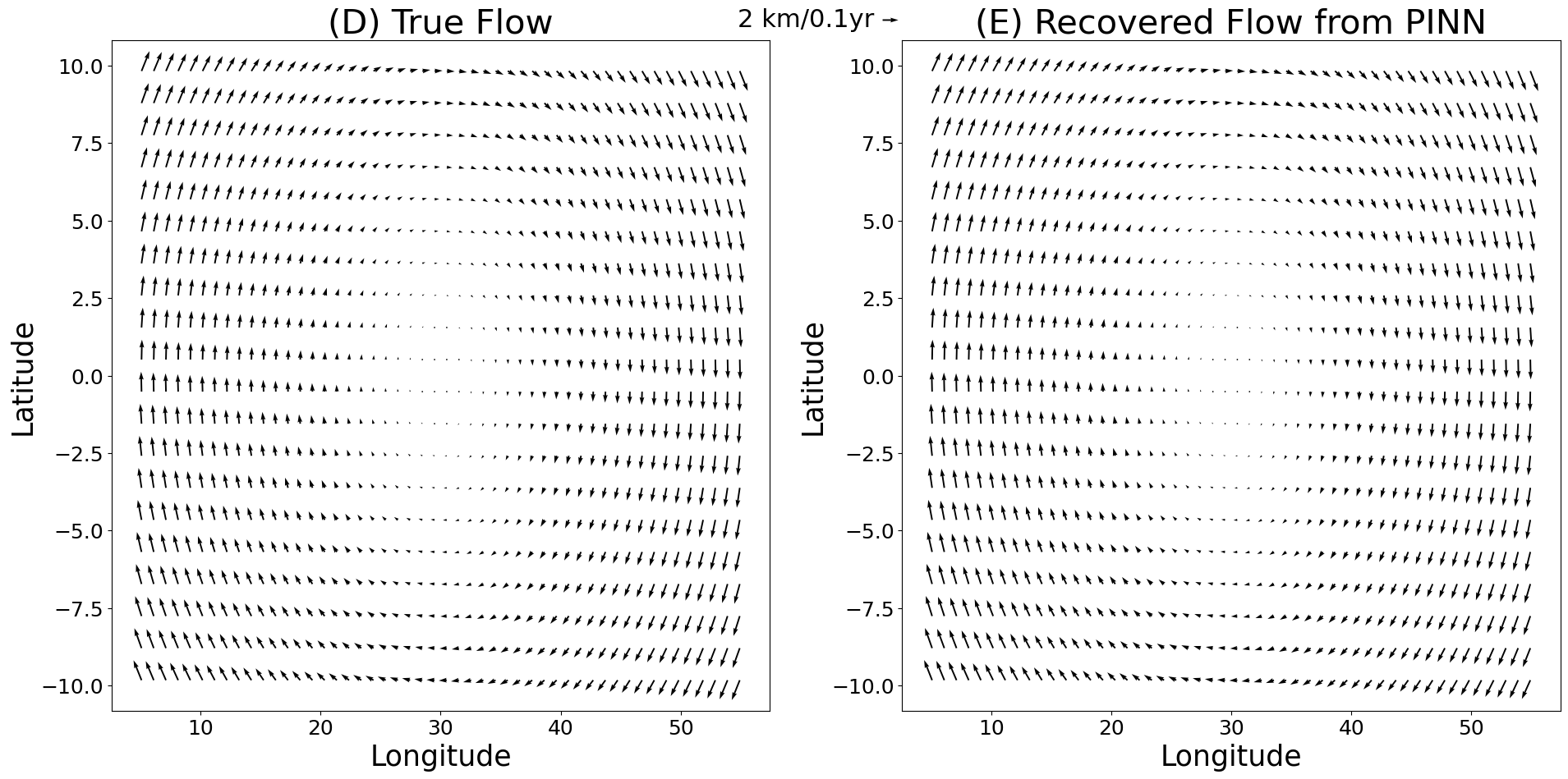}
    \caption{Example of the PINN recovery of the flow, based on a synthetic test using a Rankine Vortex (A) Input SV to the PINN, (B) Output SV from the PINN \\ (C) Residual SV (Difference between input and output)
     (D) Synthetic flow used to generate input SV (E) Output flow from the PINN. A 5$\degree$ border has been removed after training as in figure \ref{fig:box}.}
    \label{fig:syn}
\end{figure}

The models are trained with 5 different seeds, with a 5$\degree$ border removed after training as in figure \ref{fig:box}. Figure \ref{fig:syn}A shows the input SV, figure \ref{fig:syn}B shows the output SV from the PINN, and figure \ref{fig:syn}C shows the residuals between the input and output SV, all for the Rankine vortex. Figure \ref{fig:syn}D shows the synthetic Rankine flow used to generate the input SV and figure \ref{fig:syn}E shows the output flows from the PINN. Table \ref{tab:syn} shows the description and results for other synthetic flows, with the error between the true value and the output value expressed as a \% of the maximum flow value shown for each flow component, as well as the spread in the results due to the different starting seeds calculated by taking the standard deviations of the percentage errors. Each of these tests successfully recovers the synthetic flows, validating the method for large scale flows. The residuals also do not have any structures correlating to the original input, and the residuals are all small-scale. 

\begin{landscape}
\begin{table}[H]
\begin{tabular}{|l|l|l|l|l|}
\hline
\textbf{Flow Description} &
  \textbf{Flow Equations} &
  \textbf{\begin{tabular}[c]{@{}l@{}}SV RMSE\\ (\% of max value)\end{tabular}} &
  \textbf{\begin{tabular}[c]{@{}l@{}}$\pmb{u_{\theta}}$ RMSE \\(\% of max value)\end{tabular}} &
  \textbf{\begin{tabular}[c]{@{}l@{}}$\pmb{u_{\phi}}$ RMSE \\(\% of max value)\end{tabular}} \\ \hline

Eastward   & \shortstack{$u_{\theta} = 0$ \\ $u_{\phi} = \sin(\theta)$} & 0.054 $\pm 0.0074 $ & 2.4 $\pm 0.037 $  & 1.0 $\pm 0.018$  \\ \hline

Westward   &  \shortstack{$u_{\theta} = 0$ \\ $u_{\phi} = -\sin(\theta)$} & 0.063 $\pm 0.016 $ & 2.4 $\pm 0.018$  & 1.0 $\pm 0.022$  \\ \hline

Northward  & \shortstack{$u_{\theta} = \sin(\theta)$ \\ $u_{\phi} = 0 $}  & 0.065 $\pm 0.013$ & 2.3 $\pm 0.013$  & 1.0 $\pm 0.010$  \\ \hline

Southward  & \shortstack{$u_{\theta} = -\sin(\theta)$ \\ $u_{\phi} = 0 $} & 0.072 $\pm 0.018$ & 2.3 $\pm 0.029$  & 1.0 $\pm 0.02$  \\ \hline

North-East & \shortstack{$u_{\theta} = \sin(\theta)$ \\ $u_{\phi} = \sin(\theta)$} & 0.057 $\pm 0.0094$ & 4.73 $\pm 0.035$  & 2.0 $\pm 0.02$  \\ \hline

North-West & \shortstack{$u_{\theta} = \sin(\theta)$ \\ $u_{\phi} = -\sin(\theta) $} & 0.026 $\pm 0.0061$ & 0.16 $\pm 0.039$  & 0.074 $\pm 0.016$  \\ \hline

South-West & \shortstack{$u_{\theta} = -\sin(\theta)$ \\ $u_{\phi} = -\sin(\theta) $} & 0.065 $\pm 0.010$ & 4.7 $\pm 0.055$  & 2.0 $\pm 0.020$  \\ \hline

South-East & \shortstack{$u_{\theta} = -\sin(\theta)$ \\ $u_{\phi} = \sin(\theta) $} & 0.025 $\pm 0.0044$ & 0.18 $\pm 0.034$  & 0.080 $\pm 0.017$  \\ \hline

Rankine    &\shortstack{   \\ See equation \eqref{eqn:rankine}}& 0.13 $\pm 0.014$ & 0.80 $\pm 0.024$  & 0.73 $\pm 0.024$  \\ \hline

\end{tabular}
\caption{Synthetic flows used for testing. RMSE is determined by calculating the squared error at each grid point, taking the mean across the box, and then taking the square root. This is then divided by the maximum absolute value to get the percentage error. Similar results are found for flows constructed with $\cos{\theta}$.}
\label{tab:syn}
\end{table}
\end{landscape}
\newpage
\newpage
\section{Results}
\label{sec:res}
\subsection{Atlantic}
\par
\begin{figure}[H]
    \centering
    \captionsetup{justification=centering,margin=0.25cm}
    \includegraphics[scale = 0.2]{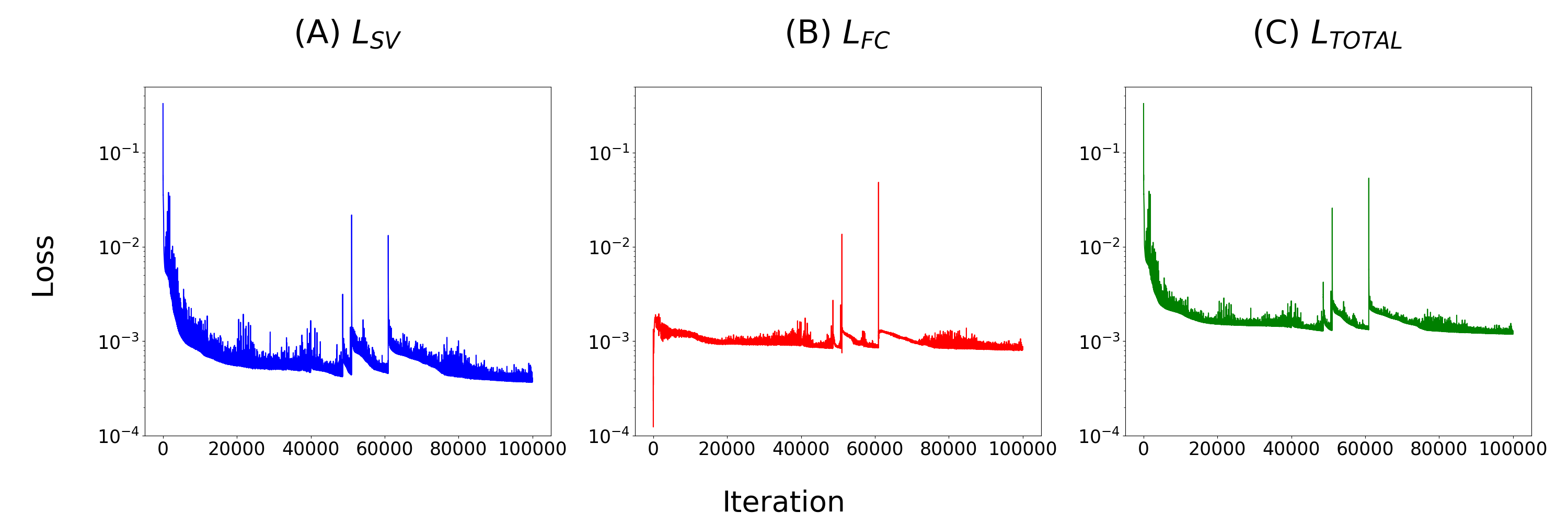}
    \caption{Loss plotted against iteration, (A) Data loss, $L_{SV}$, (B) TG Flow Constraint, $L_{FC}$, and (C) Total loss,$L_{TOTAL}$. The total loss does not decrease after 80,000 epochs, showing that the model has converged.} 
    \label{fig:loss}
\end{figure}

\begin{figure}[H]
    \centering
    \captionsetup{justification=centering,margin=0.25cm}
    \includegraphics[scale = 0.5]{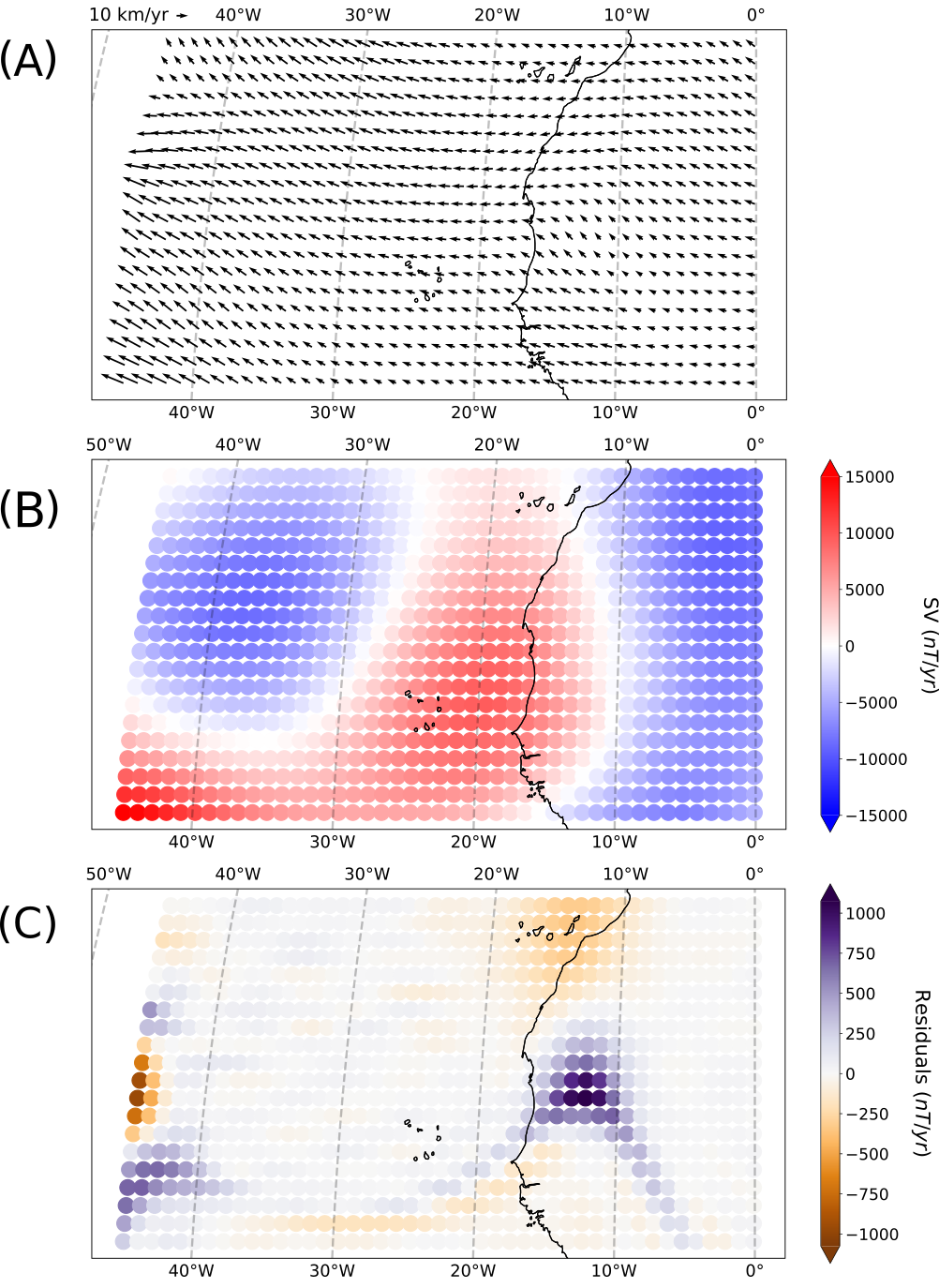}
    \caption{(A) Recovered Flow under the northern Atlantic, at 35$\degree$N to 5$\degree$N latitude, showing westward flow. \\ (B) Recovered SV \\ (C) Residuals between the SV from CHAOS-8.1 and the SV shown in (B).} 
    \label{fig:atlantic}
\end{figure}

 
Having tested the methodology on synthetic examples, we now demonstrate an application of the methodology on the region under the equatorial Atlantic, using our optimal network size. We use CHAOS-8.1 SV from 1st January 2024, training models for five different seeds, and choose the one with the lowest RMSE as the `preferred model'. Figure \ref{fig:loss}A shows $L_{SV}$, figure \ref{fig:loss}B shows $L_{FC}$, and figure \ref{fig:loss}C shows $L_{TOTAL}$, all plotted against iteration count, for one example seed. These loss curves show that the PINN has converged on a solution, and that the two loss components are of the same order of magnitude by the end of training, indicating that one loss term is not more important than the other. This vindicates our choice of a weighting factor of $\lambda = 1000$. The $L_{FC}$ term has a much smaller amplitude in the first iteration than in any other iteration, which is not the typical shape of a loss curve. This is due to the magnitude of recovered flows in the first iteration being very close to zero, thereby satisfying the TG constraint but not the SV constraint, a behaviour that is penalised by the optimiser in subsequent iterations. The inferred flows are shown in figure \ref{fig:atlantic}A, the recovered SV in \ref{fig:atlantic}B, and the residuals shown in figure \ref{fig:atlantic}C. The mean absolute error (MAE) between the SV from the CHAOS-8.1 model and the SV output by the PINN is 97nT/year (0.80\% of the maximum value of SV), and the recovered flows show the expected westward drift present in all flow inversions to date, regardless of inversion methodology (\citet{Holme_2007}, \citet{Kloss_Finlay_2019}, \citet{Whaler_Hammer_Finlay_Olsen_2022}). 

\newpage
\subsection{Global Patchwork}
\label{sec:global}
\par
In order to compare the flows recovered by the regional method with those recovered by global methods, a mosaic of regional flows is constructed from SV from 1st January 2024 using CHAOS-8.1, spanning a latitude space of within 1$\degree$ of the geographic poles. We do not include the poles in our inversion, due to the coordinate singularity there. The surface of the core mantle boundary, at a radius of 3485km, is carved up into 56 overlapping latitude-longitude boxes spanning 30$\degree$ by 55$\degree$, and 18 overlapping boxes spanning 25$\degree$ by 55$\degree$ for the boxes closest to the poles. Each of these boxes have grid points every 1$\degree$ in both dimensions.
\par
For each of the 74 boxes, a separate PINN is trained to find the flow that would generate the SV for that box only. The PINN for each box is trained for 100,000 iterations, and results obtained for 5 seeds. The model with the lowest $L_{SV}$ (SV RMSE) is taken as the `preferred model' for that box. The 5$\degree$ border is then cut off the results of each box, as in Figure \ref{fig:box}, with the result being 56 20$\degree$ by 45$\degree$ boxes, and 18 15$\degree$ by 45$\degree$ boxes. The remainder of each box are stitched together, with no continuity conditions imposed at the edges of the boxes. The results go to within 6$\degree$ of the pole.
\par
\begin{figure}[H]
    \centering
    \captionsetup{justification=centering,margin=0.25cm}
    \includegraphics[scale = 0.45]{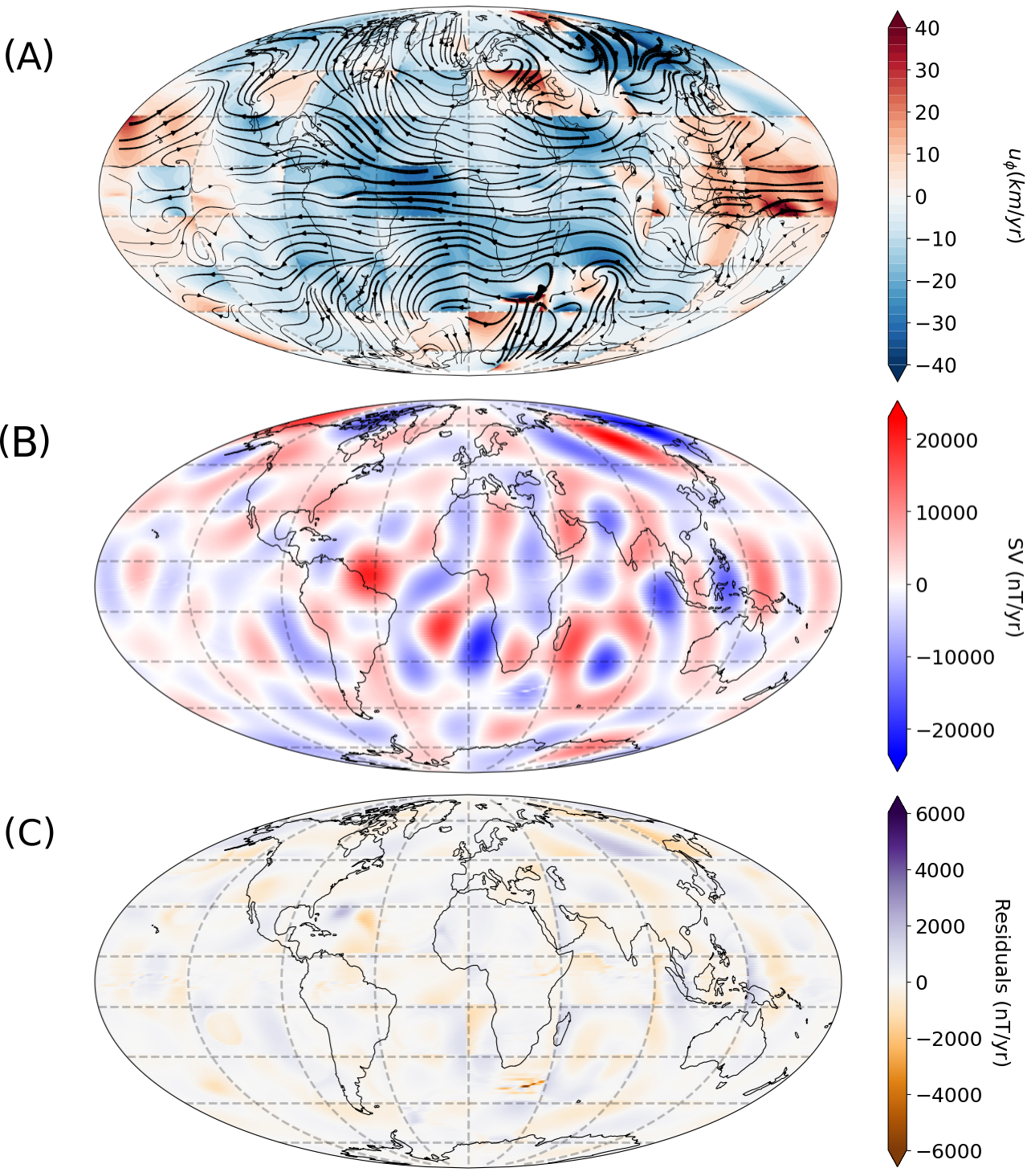}
    \caption{(A) Global patchwork of regional flows in January 2024. The thickness of the arrows indicate flow magnitude, and the arrows indicate direction. Blue indicates westward flow, and orange indicates eastward. \\\hspace{1pt}  (B) SV calculated from equation \eqref{eqn:induct}, using the recovered flows in the top panel. \\\hspace{1pt} (C) Residuals between the SV from CHAOS-8.1 and the SV shown in the middle panel. The box boundaries (after the 5$\degree$ border removal) are marked by grey dashed lines.} 
    \label{fig:global}
\end{figure}

The recovered flows, the recovered SV, and the residuals are shown in figure \ref{fig:global}.  The flows show the same large scale features as those found using global methodologies, demonstrating westward flow under the Atlantic, eastward flow under the Pacific, and the presence of the anticyclonic planetary gyre. This last feature is of particular note, as it demonstrates that this large scale feature is independent from the use of global methodologies, and is something that can be re-constructed from regional SV. The MAE between the SV from the CHAOS-8.1 model and the SV output by the PINN is 229nT/year (0.97\% of the maximum value of SV), with a standard deviation spatially of 251 nT/year. The residuals vary spatially, particularly in areas of increased SV intensity or complexity, and notably south of Africa where the maximum residual is 7578 nT/year. This high residual is consistent across seeds. Further investigation of this feature shows that the error is not consistent across time, peaking in 2024, and is not present when the TG assumption is not applied, perhaps indicating a failure of the TG assumption at this particular location. This is discussed further in section \ref{sec:discussion}.
\subsection{Equatorial Flows}
\label{sec:eq}
\par
To probe the broader flow properties of the equatorial region, the PINN methodology was applied to a 30$^\circ$ latitude by 360$^\circ$ longitude band, centred on the equator. SV values from CHAOS-8.1 were taken in one year increments from 2000.0 to 2024.0, and then five different models (one per seed) were trained for each of the 25 epochs, resulting in 125 models. For each time increment, the model from the five seeds which had the lowest RMSE between the input and output SV was chosen as the `preferred model'. We use the same network size found in section \ref{sec:son} for this elongated box, as it was found empirically that larger network sizes did not give significantly lower values of SV RMSE. Additionally, we do not impose periodicity between the longitudinal-edges of the elongated box.  
\par
\begin{figure}[H]
    \centering
    \captionsetup{justification=centering,margin=0.25cm}
    \includegraphics[scale = 0.45]{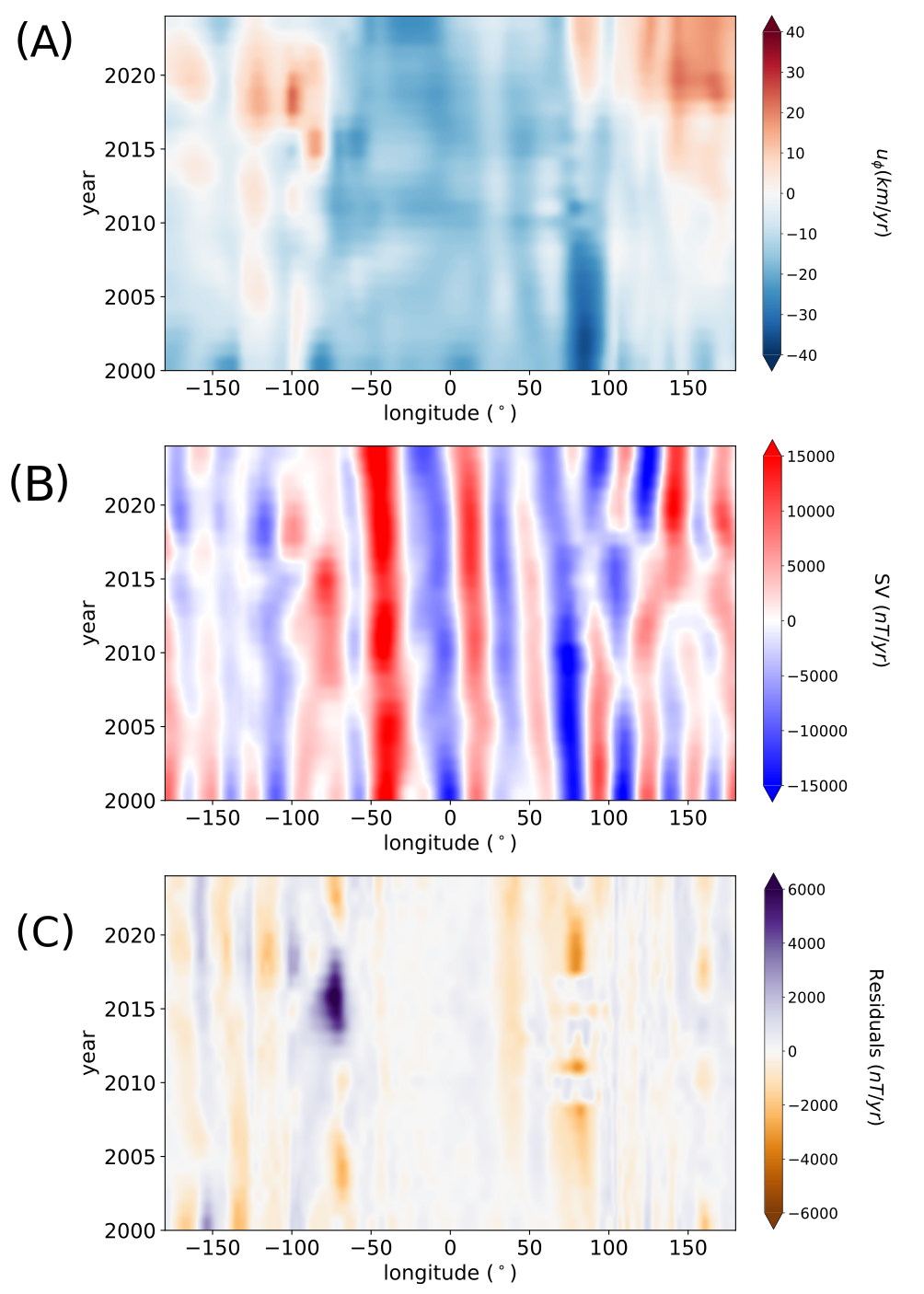}
    \caption{(A) Time-Longitude plot at the equator, spanning 2000.0 to 2024.0. Blue indicates westward flow, and orange indicates eastward. \\\hspace{1pt}  (B) SV calculated from equation \eqref{eqn:induct}, using the recovered flows in the top panel. \\\hspace{1pt} (C) Residuals between the SV from CHAOS-8.1 and the SV shown in the middle panel. }  
    \label{fig:tl}
\end{figure}

The recovered azimuthal flow at the equator was extracted from the preferred models, and displayed in a time-longitude plot, shown in figure \ref{fig:tl}A.  The area from 100$^\circ$ to 180$^\circ$ longitude shows a reversal in azimuthal flow direction, starting in 2010 and becoming more intense in 2017. This result was also found by \citet{Whaler_Hammer_Finlay_Olsen_2022}, who associated the change in azimuthal flow direction with the 2017 geomagnetic jerk. By 2024.0, this eastward flow is still present. The area -180$^\circ$ to -70$^\circ$ longitude shows features of small-scale alternating flow direction in the area under central America. This alternating pattern was previously found in \citet{Kloss_Finlay_2019} in their global flow inversion from ground- and satellite-based magnetic field measurements. Results from 2022 onwards show a recent change in the azimuthal flow direction in this area, from eastwards to westwards, which to our knowledge is a new result.
\par
Figure \ref{fig:tl}B shows the recovered SV from the PINN, and figure \ref{fig:tl}C shows the residuals between the input and output SV. The residuals are generally small, with a MAE of 434 nT/year (2.1\% of the maximum value SV), though the residuals vary slightly across longitude and time, with the largest residuals being at location of fast changes in azimuthal flow direction. Of particular note is a feature at -75$^\circ$ longitude, spanning from 2013 to 2018, where the maximum MAE is 6465 nT/year. This feature is not present when the TG assumption is not applied, once again indicating the possibility that this residual is due to a failure in the TG assumption, similar to that found in the global residuals shown in figure \ref{fig:global}C.
\newpage
 
\par
The two regions of azimuthal flow directional change indicated in figure \ref{fig:tl} --- Indonesia and Central America --- were studied in more detail. Once again, for each time increment five models were trained, and the one with the lowest RMSE was chosen to be the `preferred model'. 

\subsubsection{Indonesia}
\label{sec:ind}
\par
\begin{figure}[H]
    \centering
    \captionsetup{justification=centering,margin=0.25cm}
    \includegraphics[scale = 0.15]{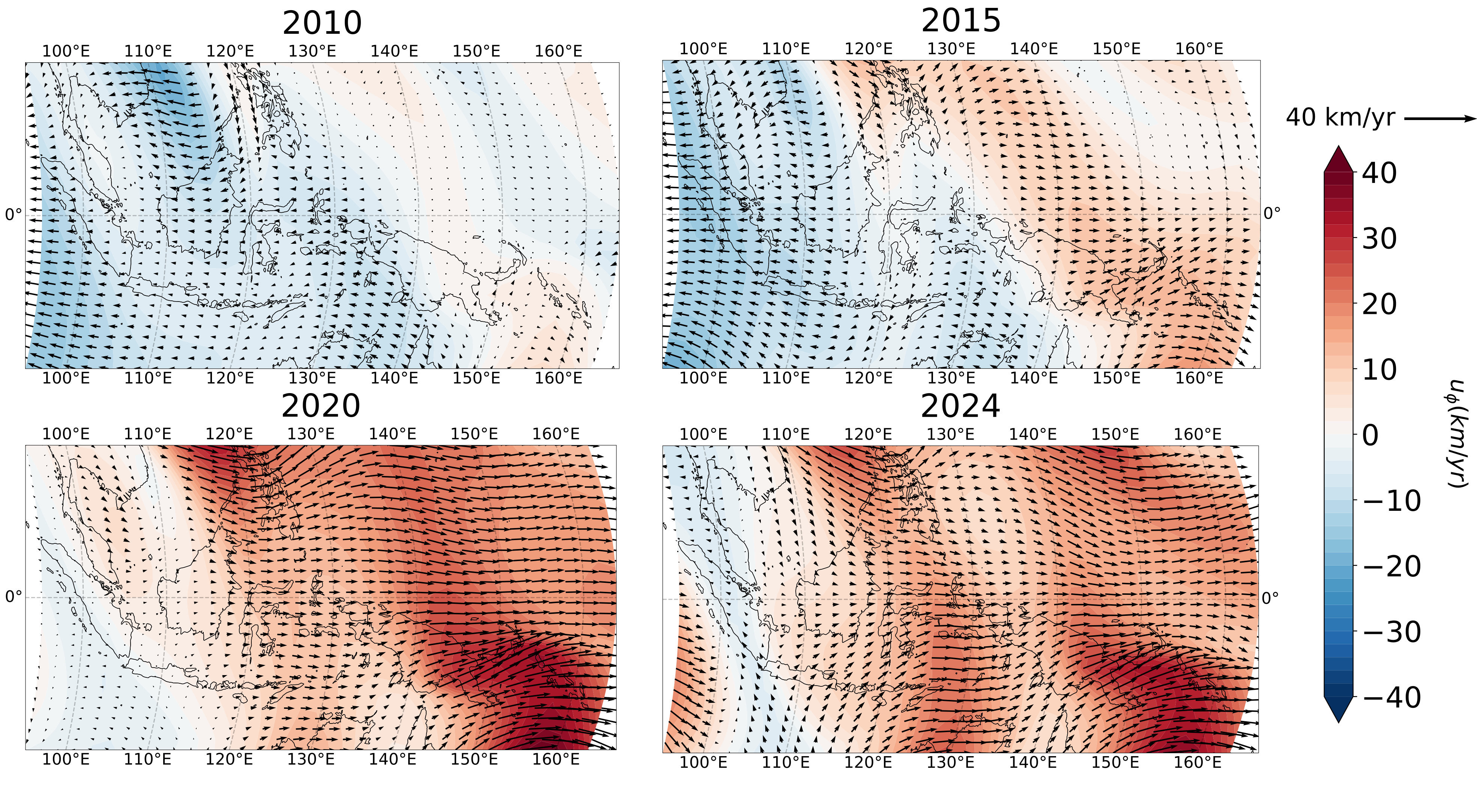}
    \caption{Changes in the horizontal flow under Indonesia during the period 2010.0 to 2024.0, at 10$\degree$ above and below the equator.} 
    \label{fig:indo}
\end{figure}
\begin{figure}[H]
    \centering
    \captionsetup{justification=centering,margin=0.25cm}
    \includegraphics[scale = 0.45]{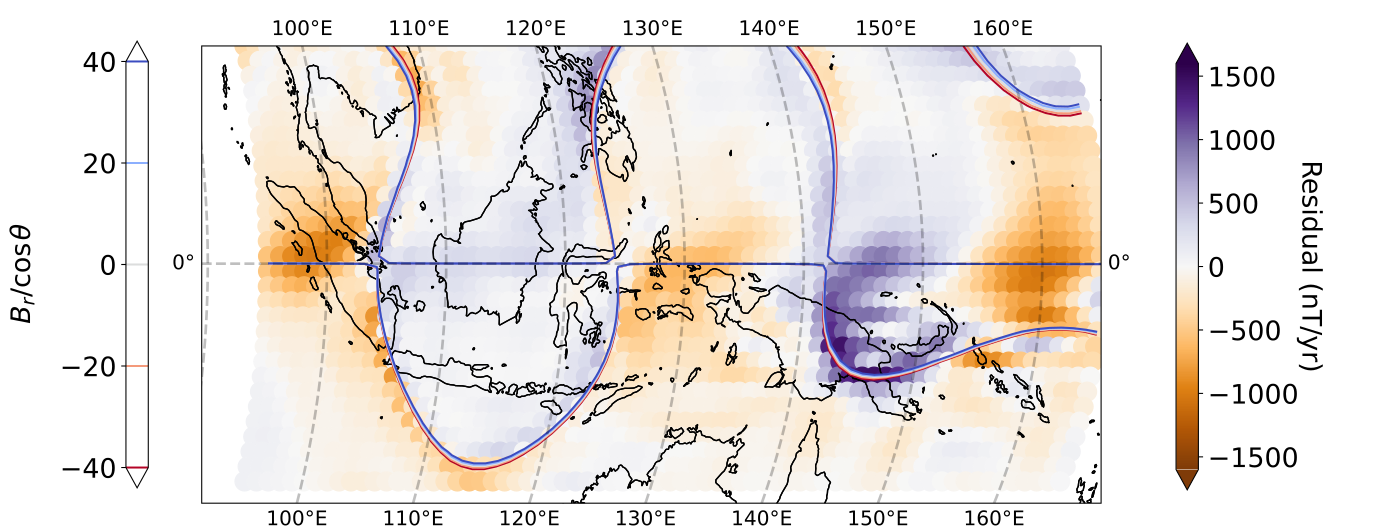}
    \caption{Residual between input SV and output SV for 2024.0, MAE of 157nT/year, with contours of $B_r/\cos{\theta}$ superimposed, truncated at $\pm$ 40 $nT/\degree$. Latitude range is 10$\degree$ above and below the equator.} 
    \label{fig:indo_sv}
\end{figure}

The recovered flows under Indonesia in the time period of 2010.0-2024.0 are shown in figure \ref{fig:indo}, and the residuals for 2024.0 shown in figure \ref{fig:indo_sv}. The flows show a predominantly westward, horizontally divergent flow structure beginning in 2010, increasing in magnitude until it becomes predominantly eastward flowing in 2020. Typical flow speeds are approximately 25 km/year, similar in magnitude to those found in \citet{Whaler_Hammer_Finlay_Olsen_2022}. Results from 2015 onwards show an increased flow magnitude at 150$^\circ$ to 165$^\circ$ longitude, which coincide with areas of increased SV residuals. These SV residuals show features that line up almost exactly with contours of $B_r/\cos{\theta}$, as shown in figure \ref{fig:indo_sv}, the lines along which the flow is ambiguous. This is perhaps explained by considering equation \eqref{eqn:tginduct}, which shows that the induced SV is proportional to $B_r/\cos{\theta}$, and we might expect a correlation between the SV residuals and $B_r/\cos{\theta}$. The average MAE across the time period of 2010.0 - 2024.0 is 171 nT/year (0.98\% of the maximum value), with a maximum of 255 nT/year in 2005 and a minimum of 120 nT/year in 2000.

\subsubsection{Eastern Pacific}
\label{sec:mexico}
\par
\begin{figure}[H]
    \centering
    \captionsetup{justification=centering,margin=0.25cm}
    \includegraphics[scale = 0.19]{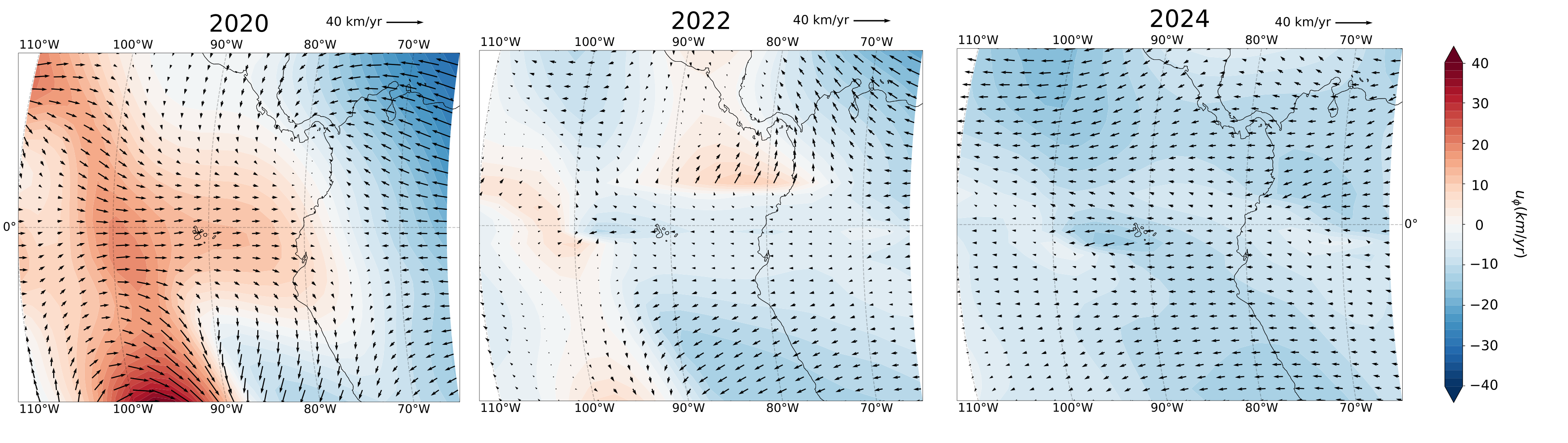}
    \caption{Changes under the Eastern Pacific during the period of 2020.0 to 2024.0, at 10$\degree$ above and below the equator.} 
    \label{fig:mex}
\end{figure}

Figure \ref{fig:mex} shows three time increments from 2020.0 to 2024.0, which were chosen to investigate the recent change in flow direction in the Eastern Pacific, between longitudes of -150$^\circ$ to -75$^\circ$. The recovered flows show a horizontally convergent flow structure in 2020, just under the coast of western South America, similar to those found in numerical simulations driven by heterogeneous CMB heat flow by \citet{Mound_Davies_2023}, although the flow is strongly time-dependent and differs only a few years later. The intensity of the eastern portion of the recovered flow decreases from 2020, before becoming entirely westward by 2024. This reversal in azimuthal flow is one of many direction changes seen in the region from 2000, with four changes in flow direction seen in the area, as presented in figure \ref{fig:tl}, and is seen in other studies such as \citet{Kloss_Finlay_2019}.  Once again, the residuals between input and output SV show features aligned with contours of $B_r/\cos{\theta}$. The average MAE across the time period of 2020.0 - 2024.0 is 201 nT/year (1.2\% of the maximum value), with a maximum of 362 nT/year in 2020 and a minimum of 75 nT/year in 2024.

\subsection{High Latitude Jet near the North Pole}
\label{sec:np}
\par
Global magnetic field models show patches of SV with alternating sign in the northern polar region, occurring close to the inner core tangent cylinder. \citet{Livermore_Hollerbach_Finlay_2017} presented an explanation for this in the form of a strengthening high latitude jet of localised flow underneath Alaska and Siberia, which forms a part of the eccentric planetary gyre. The magnitude of this jet is not well constrained, as noted in \citet{Finlay_Gillet_Aubert_Livermore_Jault_2023}; simpler, localised, models such as in \citet{Livermore_Hollerbach_Finlay_2017} show a larger flow acceleration compared to those from more complex global models such as \citet{Gillet_Gerick_Jault_Schwaiger_Aubert_Istas_2022}, which use priors to constrain non-uniqueness. It is worth noting that this choice of prior impacts the azimuthal flow speed around the tangent cylinder when using data assimilation methodologies \citep{ROGERS2025107323}. To investigate where the flows from this regional method would fit into this picture, as well as to investigate the evolution of this jet over time, the PINN method was applied to a high-latitude band spanning from 60$\degree$N to 80$\degree$N latitude, and between 0$\degree$ to 360$\degree$ longitude. Once again 5 models were trained to find the `preferred model', for increments every 5 years between 2000 and 2024, using the network size found in section \ref{sec:son}.
\par
\begin{figure}[H]
    \centering
    \captionsetup{justification=centering,margin=0.25cm}
    \includegraphics[scale = 0.3]{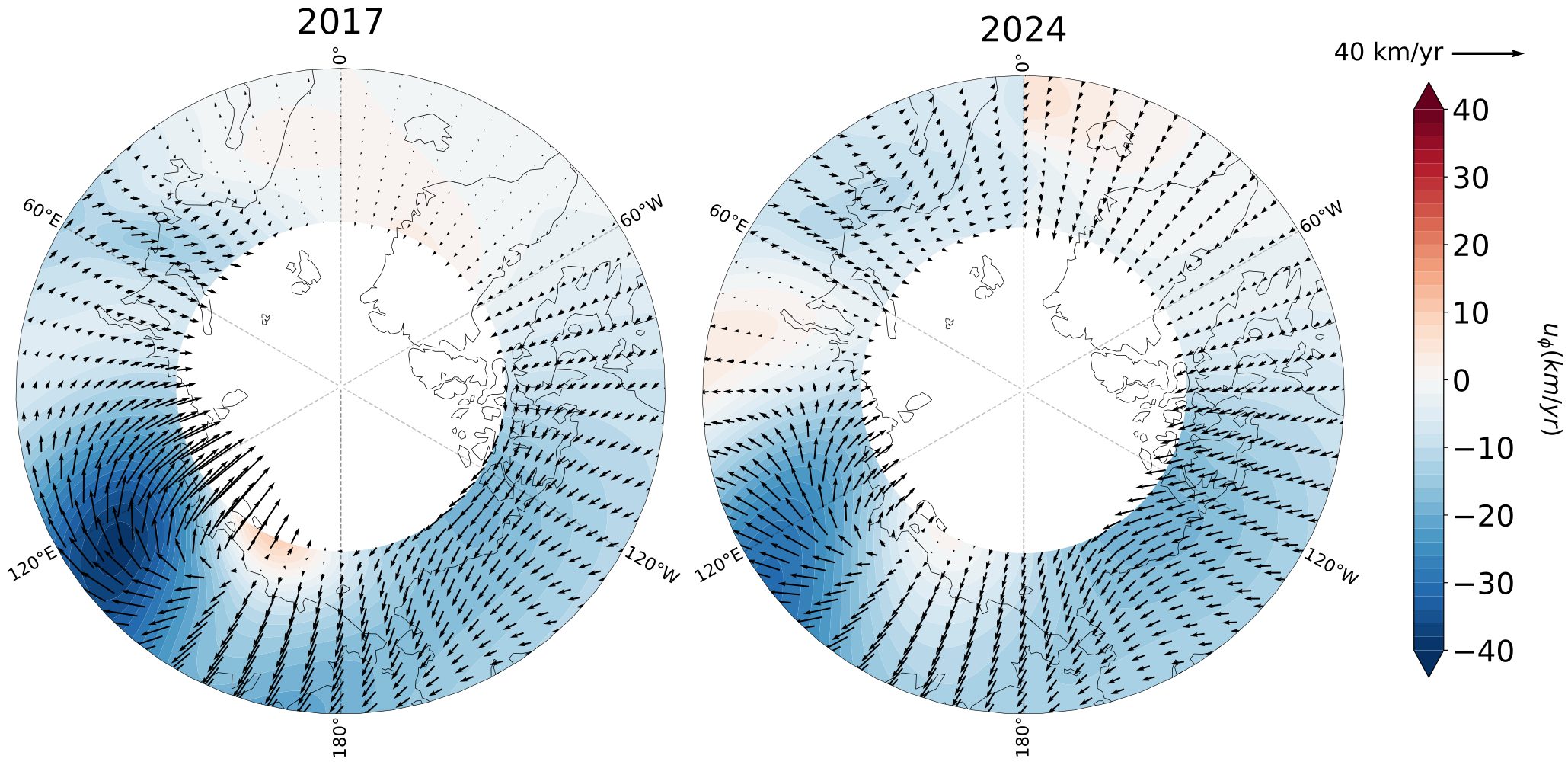}
    \caption{High Latitude jet in 2017.0 and 2024.0.} 
    \label{fig:np}
    \centering
    \captionsetup{justification=centering,margin=0.25cm}
    \includegraphics[scale = 0.5]{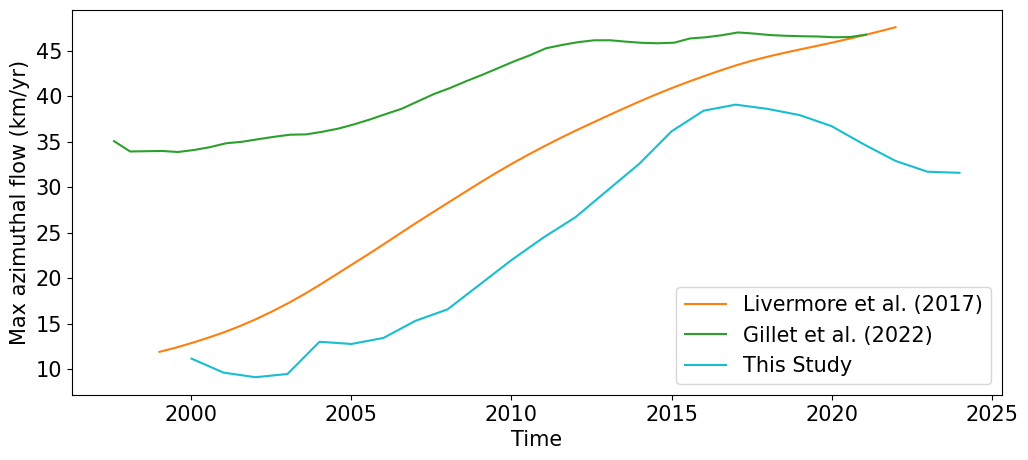}
    
    \caption{Maximum azimuthal flow over time, adapted from \citet{Finlay_Gillet_Aubert_Livermore_Jault_2023}. Results from \citet{Livermore_Hollerbach_Finlay_2017} shown in orange, results from \citet{Gillet_Gerick_Jault_Schwaiger_Aubert_Istas_2022} shown in green, this study shown in blue.}
    \label{fig:gjw}
\end{figure}

The recovered flows in 2017 and 2024 are shown in figure \ref{fig:np}.  The PINN methodology recovers the high-latitude jet, increasing in azimuthal velocity beginning in 2005, before reaching a maximum velocity of 39 km/year in 2017 and decreasing in velocity afterwards. This change in intensity may be caused by a change in the discontinuous dynamics across the inner core tangent cylinder, as suggested in \citet{Finlay_Gillet_Aubert_Livermore_Jault_2023}, and could be associated with the evolution of the planetary gyre. Figure \ref{fig:gjw} shows the maximum azimuthal flow over time from this study (blue), the localised model of \citet{Livermore_Hollerbach_Finlay_2017} (orange), and a global flow model that utilizes statistics from dynamo simulations as prior information \citep{Gillet_Gerick_Jault_Schwaiger_Aubert_Istas_2022} (green). The PINN flow results show the larger acceleration favoured by the local flow model presented in \citet{Livermore_Hollerbach_Finlay_2017}, but with a decreased amplitude compared to the other local and global flow models. We also show that the jet has been decelerating since 2017, perhaps indicating that the jet has been a transient feature that is now subsiding. This result does not change with network size or choice of seed. The average MAE across the time period of 2000.0 - 2024.0 was 107 nT/year (0.39\% of the maximum value), with a maximum of 173 nT/year in 2015 and a minimum of 49 nT/year in 2000. 

\newpage
\section{Discussion}
\label{sec:discussion}
\par
Our global mosaic model shown in figure \ref{fig:global} re-produces previously published large-scale flow structures, such as westward flow under the Atlantic, variable flow under Indonesia and Central America, and the large-scale gyre (\citet{Gillet_Gerick_Jault_Schwaiger_Aubert_Istas_2022} and \citet{Kloss_Finlay_2019}). It is important to remember that our multiple regional flow models were all trained separately and had no knowledge of the SV outside the given region, but could still produce mostly continuous large-scale features across the box edges. The reproduction of the large-scale eccentric gyre indicates this is a data-driven feature, rather than it being an artefact of the global modelling approach.  It is interesting to compare our results with the results shown in \citet{10.1093/gji/ggv552}, where the authors found that high-latitude SV can be fit locally using waves, meaning there would be no need for a gyre to connect areas at high latitude to those at mid-latitude. Our models indicate that the high-latitude flow is connected to the mid-latitude flow, suggesting that the SV at high-latitudes is a flow feature, rather than a wave feature. This does not rule out high-latitude waves, but perhaps lends weight to the idea that there is a flow component to the high-latitude SV.
\par
The time-longitude plot at the equator, shown in figure \ref{fig:tl}, demonstrates azimuthal flow direction changes happening multiple times over a period of 24 years.  There is good temporal continuation from one time-step to the next for the flow and recovered SV, despite there being no temporal regularisation applied to these PINN models. Instead, figure \ref{fig:tl} shows a series of instantaneous snapshots over time, though there is temporal smoothing applied to the CHAOS-8.1 field model that provides the input SV. The regional flow inversions shown in figures \ref{fig:indo} and \ref{fig:mex} also highlight new features, particularly the recent change in flow direction under the eastern Pacific. This behaviour could be wave-driven, similar to the waves described in \citet{Gillet_Gerick_Jault_Schwaiger_Aubert_Istas_2022}, but further work would be needed to confirm this. Recent changes are also shown in figure \ref{fig:gjw}, in the decrease in azimuthal flow intensity at the North Pole. Whether this change is due to a change in the dynamics at the inner core \citep{Finlay_Gillet_Aubert_Livermore_Jault_2023}, or due to other dynamics, is a possible area of further study.
\par
While our study presents flows modelled with the tangentially geostrophic flow assumption, this framework makes changing the flow assumption, in principle, as simple as changing a single line of code. This allows for different flow assumptions to be tested easily and quickly. Additionally, there is also no restriction on the shape of the region studied, as the loss function is a sum over a grid, which can be any shape. This provides the opportunity to study non-rectangular regions in their entirety, such as features of heterogeneity at the bottom of the mantle (e.g. Large Low-shear-Velocity Provinces, LLVPs \citep{Panton2025}). However, care must be taken when choosing a flow assumption, as the residuals presented in this study suggest there are locations and times where the TG assumption is not valid. Additionally, due to the construction method of the CHAOS-8.1 model, our local values for the main field and SV at the CMB do still depend on the observational data everywhere. For a truly regional flow inversion, a regional geomagnetic model would need to be used, which is outside the scope of this study.
\par

Regional core surface flow inversions have a wide range of potential applications, particularly in cases when the spatial distribution of data measurements is uneven, or when using paleomagnetic or historical data. The combination of this regional methodology with the SOLA (Subtractive Optimally Localised Averages, \citet{10.1093/gji/ggy515}) could be particularly productive, and may provide an opportunity for the use of historical/palaeomagnetic data for regional core flow inversions, as well as low inclination satellite orbits where data is restricted to mid- and low-latitudes, such as MSS-1 \citep{https://doi.org/10.1029/2024GL112305}.

\section{Conclusion}
\label{sec:conc}

\par
This study presents a novel methodology for inverting regional flows at the core surface, using Physics Informed Neural Networks. We validate our method on synthetic flow examples, and then construct a global flow model from a mosaic of local flow models, re-producing the large-scale gyre and therefore indicating that this is a data-driven feature. We also investigate equatorial time dependence, and present regional flow models underneath the Atlantic, Indonesia, South America and the North Pole. We highlight new features, particularly the recent change in flow direction under the Eastern Pacific and the deceleration of the high-latitude jet from 2017.  

\newpage 
\section*{Data Availability Statement}
The code produced in this study, as well as the  documentation and examples of how to use it, will be available to the community upon acceptance for publication.

The CHAOS-8 model can be accessed at: https://www.spacecenter.dk/files/magnetic-models/CHAOS-8/latest.html.

\section*{Acknowledgments} 
% THANKING ANYONE YOU NEED TO FOR DISCUSSIONS

% COMPUTING RESOURCES
This work was undertaken on ARC4, part of the High Performance Computing facilities at the University of Leeds, UK.

% INDIVIDUAL FUNDERS
NSR acknowledges support by the Leeds–York–Hull Natural Environment Research Council (NERC) Doctoral Training Partnership (DTP) Panorama (Grant NE/S007458/1). PWL was supported by Swarm DISC, funded by ESA contract no. 
4000109587. CJD and HFR acknowledge funding from Natural Environment Research Council grant NE/V010867/1. 
HFR contributed to this work while funded by the European Research Council (ERC) under the European Union’s Horizon 2020 research and innovation programme (GRACEFUL Synergy Grant agreement No 855677) and was part of Labex OSUG@2020 (ANR10 LABX56). 

\bibliographystyle{elsarticle-harv} 
\bibliography{bib.bib}
\newpage

\end{document}